\documentclass[a4paper, superscriptaddress, nofootinbib, amsmath, amssymb, amsfonts, aps, prapplied, showkeys, longbibliography, reprint]{revtex4-2}
\usepackage{graphicx}% Include figure files
\usepackage{dcolumn}% Align table columns on decimal point
\usepackage{bm}% bold math
\usepackage{xcolor}
\usepackage[english]{babel}
\usepackage[utf8]{inputenc}
\usepackage{printlen}
\usepackage[detect-all]{siunitx}
\usepackage{float}
\usepackage{upgreek}
\usepackage{comment}
\usepackage[normalem]{ulem}
\usepackage[colorinlistoftodos, color=green!40, prependcaption]{todonotes}
\usepackage{amsthm}
\usepackage{mathtools}
\usepackage{physics}
\usepackage{xcolor}
\usepackage{graphicx}
\usepackage[left=23mm,right=13mm,top=35mm,columnsep=15pt]{geometry} 
\usepackage{adjustbox}
\usepackage{placeins}
\usepackage[T1]{fontenc}
\usepackage{lipsum}
\usepackage{csquotes}
\usepackage[pdftex, pdftitle={Article}, pdfauthor={Author}]{hyperref} % For hyperlinks in the
\usepackage{epstopdf}
\usepackage{soul,xcolor}
\usepackage{svg}
\usepackage{lineno}
\usepackage{xcolor} % Подключаем пакет для работы с цветом
\begin{document}

%\linenumbers

\preprint{APS/123-QED}

\title{Thermal-noise-limited stabilization of a self-injection-locked diode laser\\ using a miniature Fabry-Pérot cavity}

\author{Alexander~K.~Vorobyev}
\email{Contact author: vorobev.ak@phystech.edu}
\affiliation{Russian Quantum Center, Skolkovo, Moscow 143025, Russia}
\affiliation{Moscow Institute of Physics and Technology, Dolgoprudny, Moscow Region 141700, Russia}
\author{Olga~V.~Khronusova}
\affiliation{Moscow Institute of Physics and Technology, Dolgoprudny, Moscow Region 141700, Russia}
\affiliation{Lebedev Physical Institute, Russian Academy of Sciences, Moscow 119991, Russia}
\author{Nikita~O.~Zhadnov}
\affiliation{Lebedev Physical Institute, Russian Academy of Sciences, Moscow 119991, Russia}
\author{Konstantin~S.~Kudeyarov}
\affiliation{Lebedev Physical Institute, Russian Academy of Sciences, Moscow 119991, Russia}
\author{Denis~S.~Kryuchkov}
\affiliation{Lebedev Physical Institute, Russian Academy of Sciences, Moscow 119991, Russia}
\author{Danila~V.~Morozov}
\affiliation{Russian Quantum Center, Skolkovo, Moscow 143025, Russia}
\affiliation{Moscow Institute of Physics and Technology, Dolgoprudny, Moscow Region 141700, Russia}
\author{Nikita~Yu.~Dmitriev}%
\affiliation{Russian Quantum Center, Skolkovo, Moscow 143025, Russia}
\author{Artem~E.~Shitikov}%
\affiliation{Russian Quantum Center, Skolkovo, Moscow 143025, Russia}
\author{Dmitry~A.~Chermoshentsev}
\affiliation{Russian Quantum Center, Skolkovo, Moscow 143025, Russia}
\affiliation{Moscow Institute of Physics and Technology, Dolgoprudny, Moscow Region 141700, Russia}
\author{Igor~A.~Bilenko}
\affiliation{Russian Quantum Center, Skolkovo, Moscow 143025, Russia}
\affiliation{Faculty of Physics, Lomonosov Moscow State University, 119991 Moscow, Russia}
\author{Nikolai~N.~Kolachevsky}
\affiliation{Lebedev Physical Institute, Russian Academy of Sciences, Moscow 119991, Russia}
\affiliation{Russian Quantum Center, Skolkovo, Moscow 143025, Russia}

\date{\today}

\begin{abstract}
We present an ultrastable laser system based on miniature core components: a self-injection-locked laser comprising a distributed-feedback laser diode and an integrated microring resonator, together with a 3-ml vacuum-gap Fabry–Pérot cavity. By carefully addressing technical noise sources, in particular residual amplitude modulation, we demonstrate a thermal-noise-limited fractional frequency instability of $5\cdot10^{-15}$ for averaging times between 1 and 10\,s, which is among the lowest reported for compact cavity-stabilized lasers. This performance is achieved with conventional multilayer mirrors and a moderate cavity finesse of $10^5$. Such stability of the output frequency, combined with the small size of the core components and a possibility to generate  a Kerr frequency comb in the very same microring resonator, provides a path toward fully integrated photonic frequency references for transportable optical clocks and low-noise microwave synthesis.
\end{abstract}

\maketitle

%%%%%%%%%%%%%%%%%%%%%%%%%%%%%%%%%%%%%%%%%%%%%%%%%%%%%%%%%%%
%                   ВВЕДЕНИЕ
%%%%%%%%%%%%%%%%%%%%%%%%%%%%%%%%%%%%%%%%%%%%%%%%%%%%%%%%%%%

\section{Introduction}

Ultrastable lasers have become a cornerstone technology for optical atomic clocks \cite{ludlow2015optical, fortier2026optical}, precision spectroscopy \cite{wan2014precision}, low-noise microwave generation \cite{kudelin2024photonic}, coherent optical communications \cite{kikuchi2015fundamentals}, radars \cite{ghelfi2011ultra}, and distributed fiber sensing \cite{lu2019distributed}. In all of these applications, the performance is directly determined by the spectral purity of the laser: a narrower linewidth and a lower fractional frequency instability translate into longer coherent interrogation times. State-of-the-art laser systems require cryogenic single-crystal silicon reference cavities \cite{lee2026frequency,chen2025laser} or ultralow-expansion (ULE) glass cavities with lengths on the order of half a meter \cite{parke2026laser,hafner20158}, allowing for fractional frequency instabilities below $10^{-16}$ and linewidths at the millihertz level. 
However, this performance comes at the price of size, complexity, and environmental isolation, which confines such systems to laboratory settings. At the same time, a growing number of field-deployable and
on-board applications --- transportable optical clocks, ranging and navigation systems, and airborne or space-based instruments --- await compact laser sources of comparable coherence \cite{grotti2018geodesy, huang2020geopotential, landerer2020extending}.

%Among various implementations of compact laser sources presented in literature, we focus on the integrated photonic solutions. 
We focus on integrated photonic solutions based on microring resonators (MR) as they offer a promising route towards miniaturization of optical systems \cite{xiang2022silicon, ye2023foundry, wan2026integrating, shekhar2024roadmapping, tian2026lithium, tran2022extending}. Indeed, integrated lasers enable the development of a new generation of efficient photonic devices that satisfy the modern concept of small size, weight, and power (SWaP) \cite{porter2023hybrid, shu2024microcomb, shen2020integrated, milvich2021integrated, pfeifle2014coherent}. %In the experimental part of work we aim to design a compact hybrid-integrated narrow-linewidth laser and achieve its long-term stabilization. Recently published papers highlight the advantages of integrated lasers for a variety of industrial applications, from fiber-optic sensors \cite{nikitin2020characterization} to optical light detection and ranging (LiDAR) systems \cite{lukashchuk2024photonic}. 
The self-injection locking (SIL) of a laser diode to a high-quality factor %$Q\sim(10^7-10^9)$ 
integrated microresonator  \cite{kondratiev2023recent, jin2021hertz, chermoshentsev2022dual} provides a direct route towards the implementation of a chip-scale source with instantaneous linewidths below 1~kHz, on par with commercial fiber lasers. The short-term as well as long-term stability of SIL-based lasers is fundamentally limited by the thermorefractive noise (TRN) of the dielectric microresonator material \cite{huang2019thermorefractive, matsko2007whispering}. While SIL allows achieving fundamentally limited narrow instantaneous linewidth, the long term stability still suffers from environmental noise. Locking the laser output frequency to an external optical etalon provides additional stabilization at intervals of several seconds \cite{hirata2014sub, guo2022chip, alnis2008subhertz, liu2024ultrastable}.

In a series of recent works \cite{kudelin2024photonic, liu2024low, cheng2025harnessing, ji2024coherent}, the miniature Fabry--P\'erot (FP) cavities (ml-volume) have successfully been used for the long-term stabilization of the SIL lasers. Locking a laser to a high-finesse cavity by the Pound--Drever--Hall (PDH) technique \cite{drever1983laser} suppresses the laser frequency noise far below the cavity linewidth, so that the residual instability is ultimately set by the Brownian thermal noise of the FP cavity spacer, substrates, and mirror coatings \cite{kessler2011fp_thermal_noise}. 
Since the fractional frequency stability scales with the relative length fluctuations $\delta L / L$ \cite{Zhadnov2018}, shortening the cavity inevitably raises the thermal noise floor. Nevertheless, even centimeter-scale vacuum-gap cavities retain thermal noise floors at the low $10^{-15}$ level \cite{mclemore2022miniaturizing} --- sufficient for the majority of field applications --- while their small size additionally reduces vibration sensitivity \cite{chen2006vibration}. Thus, the miniaturization of the reference FP cavity itself -- traditionally one of the bulkiest components of an ultrastable laser system -- is a particularly important step towards the development of compact ultrastable lasers. 
Reaching the thermal floor in practice requires careful suppression of vibration coupling \cite{chen2006vibration}, photothermal effects \cite{Yu2023,Vishnyakova2023}, and all technical noise sources in the electro-optical feedback loop, including residual amplitude modulation (RAM) \cite{shen2015systematic, zhang2016characterization}.

Our goal is to obtain the fractional frequency instability at 1 s integration time better than $10^{-14}$ in order to 
meet the requirements for transportable optical clocks, radio frequency photonic oscillators, communication, and navigation systems. In this work, we  demonstrate an ultrastable laser system containing a compact laser diode self-injection-locked to an integrated $\mathrm{Si_3N_4}$ MR and a miniature ($V < 3$~ml) vacuum-gap FP cavity serving as a long-term reference. In contrast to previously presented results based on coil resonators \cite{guo2022chip,loh2025optical}, we employ a ring microresonator, which allows to obtain both linear SIL operation and nonlinear regimes involving frequency comb generation. Additionally, in comparison with previously reported techniques \cite{guo2022chip, kudelin2024photonic}, we actuate the SIL laser output using laser diode current control, rather than commonly used acousto-optic frequency actuation. We demonstrate SIL laser frequency stabilization both in linear \cite{kondratiev2017self} and nonlinear SIL \cite{voloshin2021dynamics} regimes (accompanied by frequency comb generation). The system reaches the cavity thermal noise limit with a fractional frequency instability of $5\cdot10^{-15}$ at averaging times of 1--10~s, which is one of the lowest values reported for cavity-stabilized lasers with ml-scale etalons. Together with the compatibility of ring microresonators with comb generation, our results establish a realistic path toward a compact, thermal-noise-limited photonic frequency reference for transportable optical clocks and low-noise microwave photonics.

%%%%%%%%%%%%%%%%%%%%%%%%%%%%%%%%%%%%%%%%%%%%%%%%%%%%%%%%%%%
%                   ЭКСПЕРИМЕНТ
%%%%%%%%%%%%%%%%%%%%%%%%%%%%%%%%%%%%%%%%%%%%%%%%%%%%%%%%%%%

\section{Stabilization concept}
The system consists of a distributed-feedback (DFB) laser diode, which frequency is self-injection-locked to an eigenmode of an integrated $\mathrm{Si_3N_4}$ MR, and a miniature FP cavity that provides a long-term optical frequency reference. First, we employ the DFB laser diode butt-coupled to the integrated $\mathrm{Si_3N_4}$ MR \cite{raja2019electrically} as a hybrid-integrated SIL laser. Next, we lock the frequency of the SIL laser to the external reference cavity using PDH technique \cite{drever1983laser, black2001introduction}.  The resulting error signal is applied to the DFB diode current, which acts directly on the SIL laser frequency and eliminates the need for an additional acousto-optic frequency actuator. In the linear SIL regime, the laser frequency varies almost linearly with the diode current within the locking range, providing an appropriate actuator for the PDH servo \cite{shitikov2022optimization,kondratiev2023recent,kondratiev2017self}. 

The complete stabilization system comprises three feedback branches derived from the FP cavity signal. The first branch controls the laser diode current and stabilizes the optical frequency. The second stabilizes the optical power incident on the reference cavity, while the third actively suppresses RAM. Together, these loops stabilize the relevant degrees of freedom that can perturb the PDH lock and thereby enable operation close to the thermal-noise limit of the miniature FP cavity.

%Additionally, at the optimum, the locking ranges for increasing and decreasing frequency scans become approximately equal, corresponding to the maximum stabilization coefficient (for details see Refs. \cite{shitikov2022optimization, kondratiev2023recent, kondratiev2017self}). 
%Typically, after adjustment we obtain the transmission signal shown by green line in Fig.~\ref{transmission}. 
%

\section {Experimental setup}

\subsection{Laser diode self-injection locking}

\begin{figure*}[hbtp!]
    \centering
    \includegraphics[width=\linewidth]{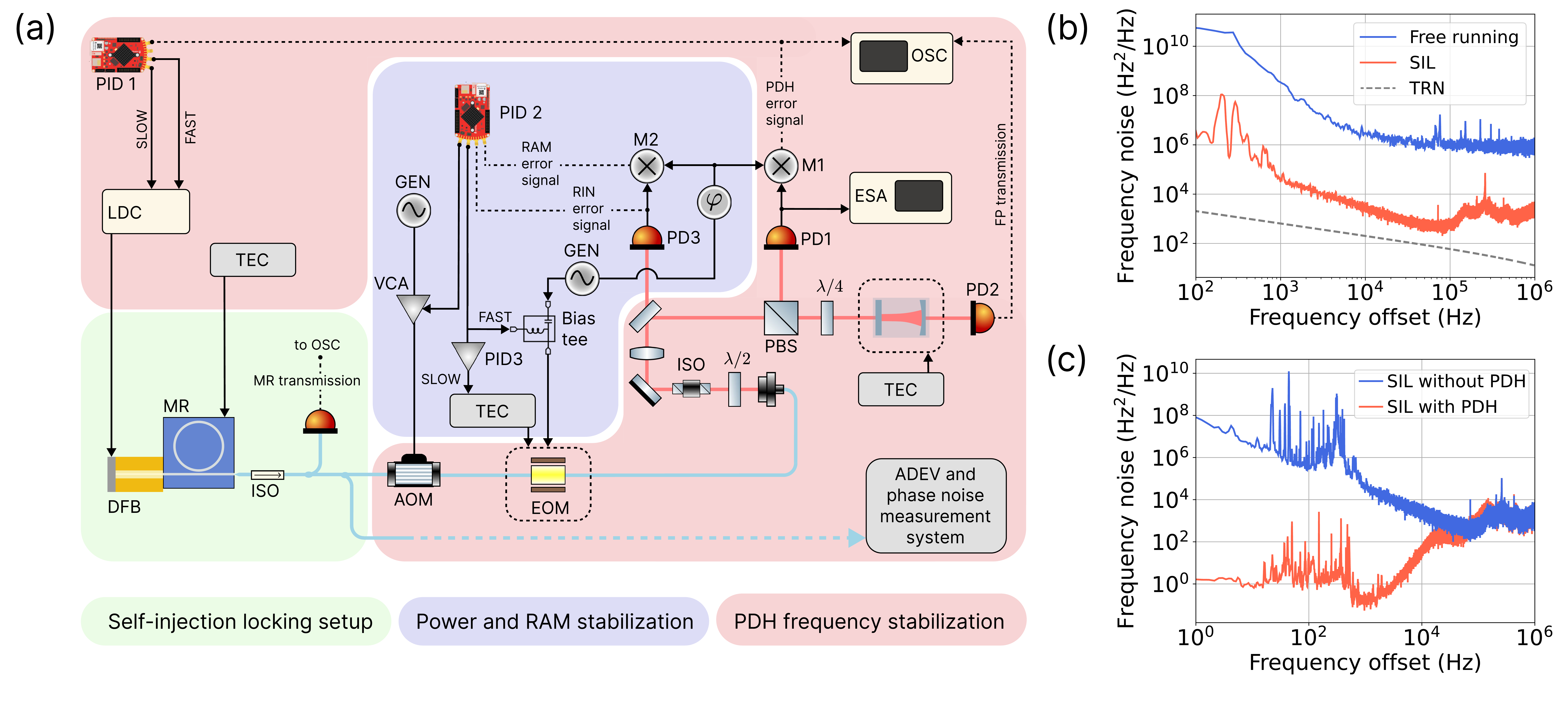}
    \caption{\textbf{(a)} Experimental setup for PDH locking of the SIL laser to a FP cavity and characterization of its performance. LDC - laser diode controller, MR - microring resonator, DFB - distributed feedback laser, TEC - temperature controller, ISO - optical isolator, PID - proportional-integral-derivative controller, GEN - radiofrequency generator, VCA - voltage-controlled radiofrequency amplifier, AOM - acousto-optic modulator, EOM - electro-optic modulator, M1 and M2 - radiofrequency mixers, PD - photodetector, PBS - polarizing beamsplitter cube, ESA - electrical radiofrequency spectrum analyzer, OSC - oscilloscope; RAM - residual amplitude modulation, RIN - relative intensity noise. The labels "slow" and "fast" refer to two different feedback loops within the same control subsystem (e.g. RAM stabilization system). \textbf{(b)} Experimentally measured frequency noise PSD for the DFB laser in the free-running (without optical feedback from MR) regime and in the SIL regime. Gray dashed line shows the analytically calculated thermorefractive noise (TRN) level for the MR under study in the infinite heat bath approximation \cite{huang2019thermorefractive}. \textbf{(c)} Experimentally measured frequency noise PSD for SIL laser with and without PDH locking to the FP cavity.}
    \label{pdh_exp_setup}
\end{figure*}

The experimental setup is presented in Fig.~\ref{pdh_exp_setup}(a). We use a 1545 nm DFB laser with up to 150 mW output power as a pump. The laser is butt-coupled to a silicon nitride photonic chip with a microresonator with total Q-factor $Q=8.2\cdot10^6$, and a free spectral range of $\mathrm{FSR}=21.3\,$GHz; the characterization details are presented in Methods B. The temperature of the chip is stabilized with a temperature controller (TEC) at $T_0=25.0\ ^o\mathrm{C}$ with 1 mK stability. The distance between the chip and the laser allows for aligning the backscattering phase from the microresonator and can be adjusted with piezo translation. For the linear SIL regime, the DFB laser output power is 40 mW, while after the silicon nitride chip, the output in SIL regime is 1.5 mW.

\subsection{Pound-Drever-Hall setup}

The reference FP cavity is made of ULE glass and has a length of $L=2\ \mathrm{cm}$, a finesse of $\mathcal{F}=105000$, and a quality factor of $Q=2.7\cdot10^9$; it is housed in a vacuum chamber and is temperature-stabilized at its zero-coefficient-of-thermal-expansion temperature of $T_0=27.35\ ^\circ\mathrm{C}$. The detailed characterization of the FP cavity is presented in Methods D.

Phase modulation of the laser beam is realized using a temperature-stabilized fiber-coupled electro-optic modulator (EOM) with a built-in polarizer. The EOM is driven by a 9.1 MHz signal supplied by an RF generator. After passing through the EOM, the optical beam is collimated into the free-space section of the setup, which is located on the optical table together with  the FP cavity with no active vibration isolation. The optical beam, with a power of 30 $\mathrm{\upmu W}$, is mode-matched to the FP using a lens.

The light reflected from the cavity is detected by a photodetector PD1. Its output signal is amplified and demodulated on a phase detector using the second output of the RF generator that provides the modulation signal. The relative phase between the modulation and demodulation signals is adjusted to maximize the slope of the PDH error signal. The resulting error signal is processed by a digital PII$^2$ controller implemented on a STEMLab 125-14 single-board computer (PID 1). The controller output signal modulates the current of the laser diode through a custom-designed shunt circuit placed after the laser diode controller (LDC) Toptica DCC 110. The feedback bandwidth is approximately 100 kHz and is limited primarily by electronics latency. We also send the additionally integrated output of the controller to the slow, wide-range current modulation input of the laser diode controller to improve robustness against long-term frequency drifts and large perturbations.

\subsection{Power stabilization}
The optical power is stabilized using a voltage-controlled amplifier (VCA) that regulates the RF drive power applied to the fiber-coupled acousto-optic modulator (AOM), installed prior to the EOM. Note that we employ the AOM exclusively for optical power stabilization, whereas frequency stabilization is achieved by controlling the DFB laser diode current. Consequently, in the experimental approach used, the AOM may be replaced by a more compact element, such as an integrated intensity modulator.

\subsection{Stabilization of residual amplitude modulation}

A detailed characterization of the stabilization system reveals that fluctuations of the RAM \cite{shen2015systematic} constitute the dominant source of frequency instability at long averaging times. Passive RAM suppression is partially achieved by introducing optical isolators to eliminate parasitic etalons; however, this results in only a minor improvement. Although the fiber-coupled EOM offers a compact design and a low half-wave voltage, it has poor long-term stability of the alignment between the input polarization and the crystal axes, leading to a fractional frequency instability of approximately $10^{-13}$ for averaging times $\tau \geq 1$~s.

To actively stabilize the RAM level, we split off approximately half of the optical power before the cavity and direct it to a photodetector PD3 identical to that used for PDH detection. One branch of the photodetector output is demodulated to extract the RAM signal, while the second branch is used to monitor and stabilize the optical power incident on the cavity. Both signals are processed by a second STEMLab 125-14 board (PID~2), where independent digital PI controllers generate the corresponding feedback signals.

The RAM feedback is applied to the EOM through a bias tee, which combines the RF drive signal with the DC-coupled feedback. The integrator in this feedback loop gradually saturates over time. To prevent saturation, we slowly drive its output back toward zero by adjusting the setpoint of the EOM temperature control system. Indeed, active RAM stabilization is a key feature of the developed experimental technique, enabling the reduction of the fractional frequency instability level below $10^{-14}$ over averaging times from 0.1 to 10 s. For comparison, we show the experimentally measured Allan deviation with and without active RAM stabilization in Fig.~\ref{ram noise} (see Appendix E).

\section{Results}

%\subsection{Laser in linear SIL regime}

To characterize the performance of the stabilized laser, we generate a beat-note between its output and the corresponding spectral component of an optical frequency comb manufactured by Avesta. Since the source of the optical frequency comb is located in a separate laboratory, the stabilized SIL laser radiation is transferred through an optical fiber link with active fiber-noise cancellation. The optical frequency comb is referenced to a 1140 nm clock laser used for thulium optical clocks \cite{golovizin2024tm}. The 1140 nm laser exhibits a fractional frequency instability of approximately $2 \cdot 10^{-15}$ at an averaging time of 1 s, which is lower than our FP cavity thermal noise limit ($5 \cdot 10^{-15}$).

The frequency noise suppression of the SIL laser compared to the free-running laser is presented in Fig.~\ref{pdh_exp_setup}(b). The instantaneous linewidth of the laser becomes close to the thermo-refractive noise \cite{huang2019thermorefractive}, see the gray dashed line in Fig.~\ref{pdh_exp_setup}(b). Experimentally measured frequency noise power spectral density (PSD) for the SIL laser with and without PDH locking to the FP cavity is presented in Fig.~\ref{pdh_exp_setup} (c). The PDH stabilization effectively suppresses the frequency noise at Fourier frequencies below $10$ kHz within the feedback loop bandwidth, providing more than 75 dB of noise reduction at 1 Hz. Sharp resonances observed between 10 and 1000 Hz are attributed to mechanical vibrations of the cavity, as the experimental setup does not employ active vibration isolation.

\begin{figure}[htbp!]
    \centering
    \includegraphics[width=\linewidth]{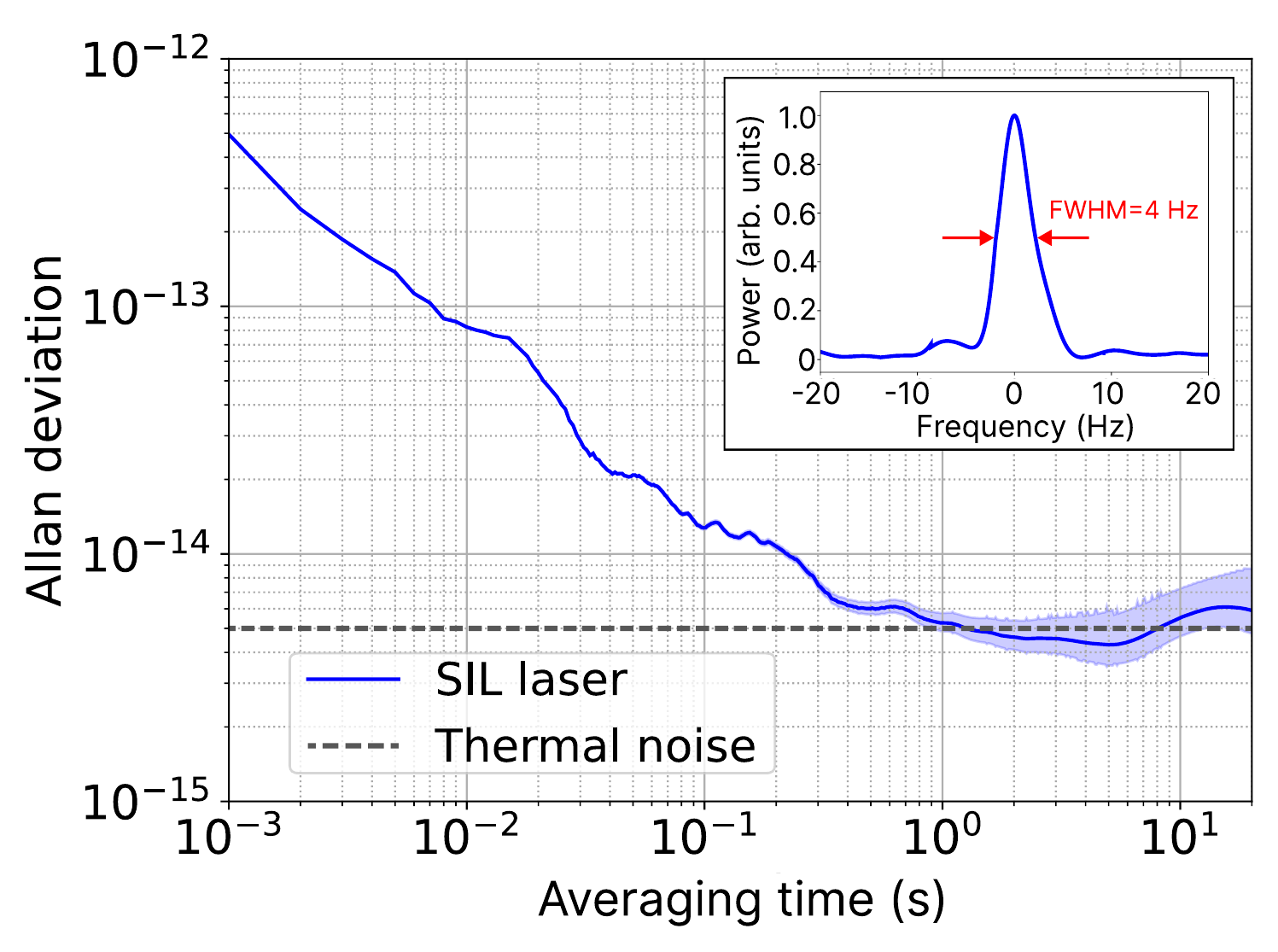}
    \caption{Main figure: experimentally measured fractional frequency instability of the SIL laser referenced to the FP cavity. Transparent color shows the confidence interval of the measured instability. Inset: measured beatnote spectrum between the SIL laser and the stabilized frequency comb. Laser spectral line FWHM is estimated to be 4 Hz, which is mostly limited by the instrument RBW = 3 Hz. See the main text for details.}
    \label{allan sil laser}
\end{figure}

To characterize the long-term frequency stability while accounting for the instability of the 1140 nm reference laser, we employ the three-cornered-hat method \cite{gray1974three}. This technique allows the individual frequency instabilities of the lasers to be estimated from pairwise beat measurements. Details of the measurement procedure are provided in Appendix C. For this analysis, two ultrastable lasers are simultaneously compared with the comb: the stabilized SIL laser developed in this work and the 871 nm laser used for ytterbium-ion spectroscopy in a neighboring laboratory \cite{galstyan2025yblaser}. We record the two beat signals with a frequency counter, remove their linear frequency drifts, and calculate the individual Allan deviation of each laser. The resulting fractional frequency instability of the SIL laser is $5 \cdot 10^{-15}$ for averaging times between 1 and 10 s (see Fig.~\ref{allan sil laser}), consistent with the thermal-noise limit of the FP cavity.

For the direct linewidth measurements, we record the beatnote spectrum  between the SIL laser and the stabilized frequency comb using a spectrum analyzer with a resolution bandwidth (RBW) of 3 Hz. The measured spectrum is shown in Fig.~\ref{allan sil laser}, inset. From these data, we estimate the FWHM to be 4 Hz, which is mostly limited by the instrument RBW.

% \begin{figure}[h]
% \centering
% \includegraphics[width=\linewidth]{figs/fwhm_v2.png}
% \caption{SIL laser beatnote with the frequency comb stabilized to the reference ultrastable laser.}
% \label{lorenztian_fit}
% \end{figure}

\begin{figure}[h]
    \centering
    \includegraphics[width=\linewidth]{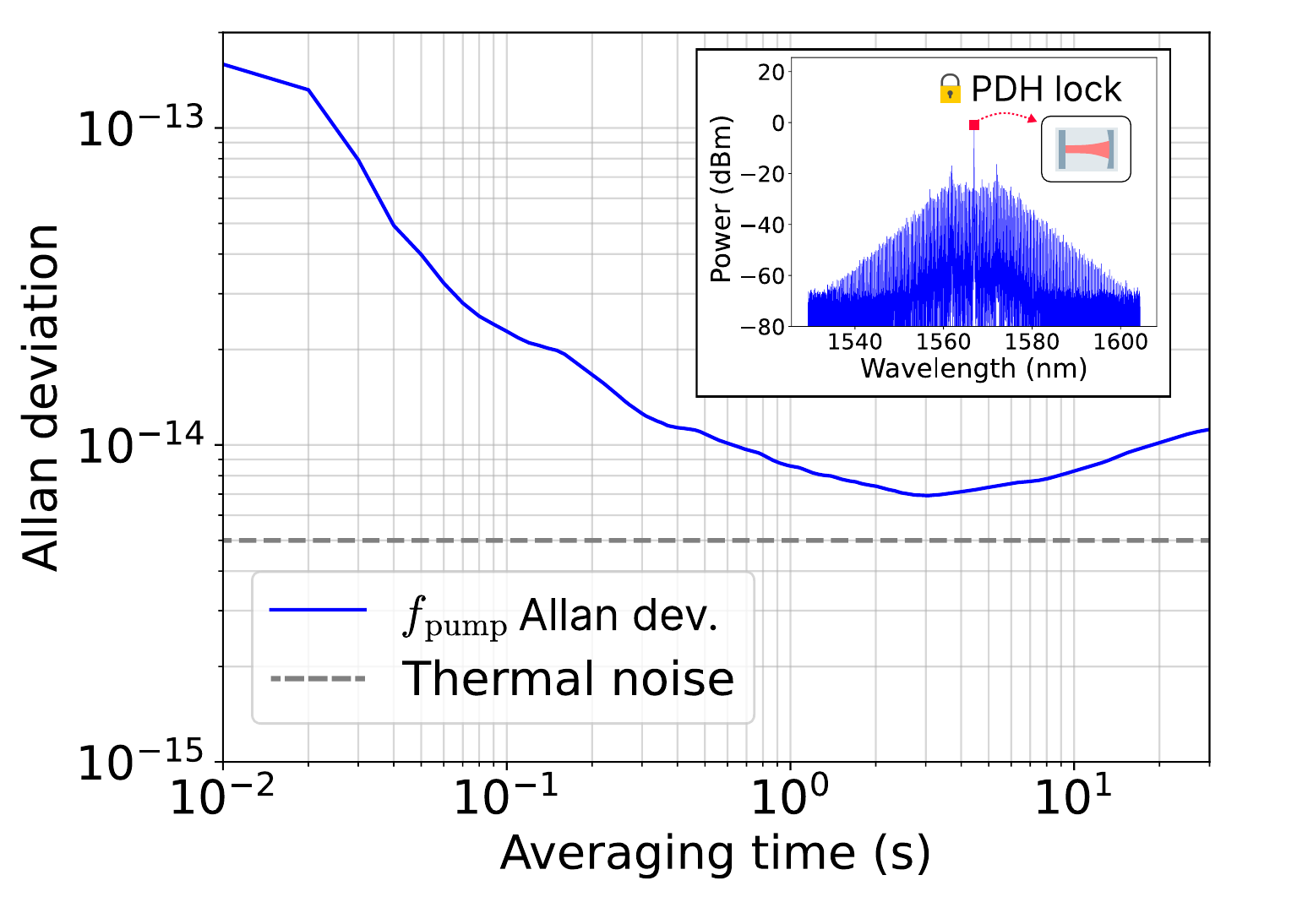}
    \caption{Allan deviation of the pump frequency $f_\mathrm{pump}=191.45$ THz in the Kerr comb generation regime (see comb spectrum in the inset). The comb repetition rate $f_\mathrm{rep}=30.6$ GHz.}
    \label{comb_and_freq_noise}
\end{figure}

Besides single-frequency operation, we examine whether stabilization can be maintained in the Kerr comb generation regime (nonlinear SIL regime \cite{voloshin2021dynamics}). If successful, this approach would enable stabilization of the central comb line without requiring an additional CW laser and a phase-locked loop, which are typically used for this purpose. We employ a different  MR with anomalous group velocity dispersion $\beta_2=-142.7\ \mathrm{ps^2/km}$, free-spectral range $\mathrm{FSR}=30.6\ \mathrm{GHz}$, and quality factor $Q=6.3\cdot10^6$. At a higher DFB current value ($P_\mathrm{out} = 100$ mW), we excite a soliton frequency comb in the MR. The corresponding soliton spectrum is presented in the inset of Fig.~\ref{comb_and_freq_noise}. Using the same experimental procedure as for the locking of a single-frequency SIL laser, we achieve the locking of the comb pump tooth ($f_\mathrm{pump}=191.45$ THz) to the FP cavity eigenmode. As a result, the measured Allan deviation of $f_\mathrm{pump}$ is below $1\cdot10^{-14}$ for averaging times of 1--20~s (see Fig.~\ref{comb_and_freq_noise}), approaching the stability achieved in the linear SIL regime.

In fact, following the same procedure with a platicon microcomb excited in the MR with normal GVD (used in the main study), we find that the pump frequency $f_\mathrm{pump}$ exhibits higher fractional frequency instability($\approx1\cdot10^{-13}$ at 1 s), compared to the soliton microcomb ($\delta f/f<1\cdot10^{-14}$ at 1--20 s). The corresponding platicon spectrum is presented in Appendix H. This reduction in stability may be related to the influence of the comb sidebands nearest to the pump, as no optical band-pass filter is used to suppress them. Indeed, the nearest soliton sidebands lie 25 dB below the central comb line (inset in Fig.~\ref{comb_and_freq_noise}), while the nearest sidebands of the platicon microcomb are only 8 dB below the pump (see Appendix H). However, this phenomenon requires further investigation.

The locking of the pump tooth provides one of the two feedback loops usually required for full frequency stabilization of a comb. Preliminary measurements show that the instability of the comb teeth increases with their spectral distance from the pump line. This behavior is expected since the frequency fluctuations of the $n$-th tooth, $\delta\nu_n = \delta\nu_{\mathrm{p}} + n\,\delta f_{\mathrm{rep}}$, are dominated by the free-running repetition rate when only the pump is stabilized. Implementing optical frequency division techniques \cite{kudelin2024photonic, ji2025dispersive,diakonov2025hybrid} with a second stabilization loop acting on $f_{\mathrm{rep}}$ will therefore be required to transfer the cavity stability to the entire comb spectrum and to the microwave domain.
Though the careful examination of platicon and soliton SIL microcombs locking is the subject of our future research, the demonstrated proof-of-concept experiment allows us to rely on the designed stabilization technique in the further design of optical frequency division schemes.

%However, this is more than an order of magnitude worse than our best result obtained in linear SIL regime (single-frequency SIL laser locking). We associate this stability reduction with the additional noise caused by the interaction of all comb modes with the FP cavity, but identifying the exact cause is the subject of further research. 

% \begin{figure}[H]
%     \centering
%     \includegraphics[width=\linewidth]{figs/Изображение JPEG-49C8-A87B-30-0.jpeg}
%     \caption{(a) Allan deviation of the stabilized pump line in case of Kerr-comb generation. The specrum of the comb is presented in the inset. (b) Frequency noise of the microcomb repetition rate ($f_\mathrm{rep}=21.3$ GHz) for the cases of PDH-locked and unlocked comb pump tooth ($f_\mathrm{pump}=193.5$ THz).}
%     \label{comb_and_freq_noise}
% \end{figure}

% In the microwave domain we manage to observe the influence of pump tooth locking on the noise of the microcomb repetition rate $f_\mathrm{rep}=21.3$ GHz. Indeed, due to the inherent properties of Kerr microcombs, the pump frequency noise typically contributes to the microcomb repetition rate noise \cite{matsko2015noise, liang2015high, savchenkov2008phase, lei2022optical, taheri2025dynamics, taheri2017optical}. We find that the locking of the pump tooth reduces the frequency noise of the microcomb repetition rate by up to 20 dB at Fourier offset frequencies below 1 kHz, as shown in Fig.~\ref{comb_and_freq_noise}(b). 

\section{Discussion}
%АНАЛИЗИРУЕМ ВКЛАД РАЗЛИЧНЫХ ШУМОВ В НАШУ СИСТЕМУ И ПРЕДЛАГАЕМ ВАРИАНТЫ, КАК МОЖНО УЛУЧШИТЬ ШУМЫ НА ТЕХ ИЛИ ИНЫХ ВРЕМЕНАХ
%-можно сделать без крутых зеркал, с не очень высокой резкостью
%-достигли хороших результатов за счет тщательной настройки электроники
%-упомянуть про гребенку чтобы усилить статью, привязывается даже в режиме гребенки с плохой стабильностью и мы можем дальше изучать, как повысить стабильность, обосновать, зачем это делать
%-сделаем гребенку на 1762 и будем использовать как часовой лазер 
%перспектива на будущее: меньше тепловой шум за счет кварцевых подложек зеркал, меньше вакуумная камера и система стабилизации, лучше финес, лучше устойчивость к вибрациям
%-подчеркнуть что используем треуголку и наша характеризация ОЧЕНЬ ДОСТОВЕРНАЯ
%-в атоме Ti часовой переход на 1550
%-сравнить улиточные резонаторы и кольцевые
%-подчеркнуть, что достигнут фундаментальный предел
%
%-меньше система
%-выше стабильность

We achieve a telecom band stabilized laser with a fractional frequency instability of $5\cdot10^{-15}$ at averaging times of 1--10 s. This value represents the intrinsic instability and is independent of the instability of the ultrastable reference laser, owing to the use of the three-cornered hat method. 
%This performance is attained through careful optimization of the laser frequency stabilization system and effective suppression of various noise sources. Notably, this result is obtained using a FP cavity with conventional multilayer dielectric mirrors and a moderate finesse. In contrast, the previous best stability reported for integrated lasers has been achieved using microfabricated mirrors with a finesse approximately nine times higher \cite{jin2022micro,mclemore2022miniaturizing}, which require considerably more advanced fabrication technology.
The present performance is fundamentally limited by the thermal noise of the reference cavity at averaging times between 1 and 10 s. Consequently, further improvement requires reducing the cavity thermal noise. However, preserving the compactness of the system is essential; therefore, neither increasing the cavity length nor employing cryogenic operation is a viable option. The most practical approach is to use fused-silica mirror substrates, which are expected to reduce the thermal noise by approximately a factor of two.

Environmental temperature fluctuations affecting both the cavity and residual amplitude modulation are expected to become significant only at averaging times of several hundred seconds. In contrast, improving the performance at averaging times below 1 s is more urgent since it has the greatest impact on the laser linewidth.
\begin{figure*}[t]
    \centering
    \includegraphics[width=0.75\textwidth]{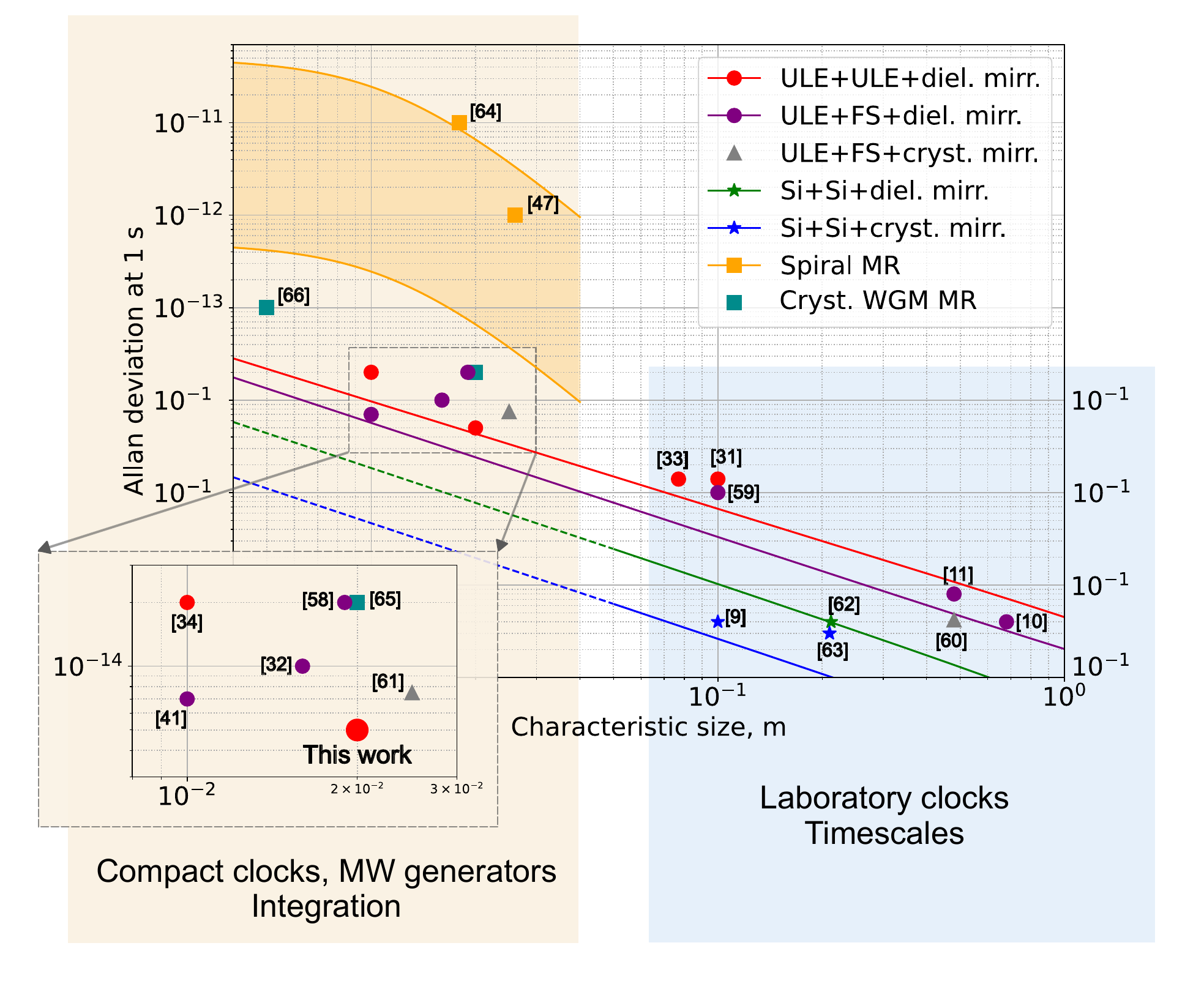}
    \caption{Allan deviation (averaging time = 1 s) for different laser systems as a function of the characteristic size of the systems: length for Fabry–Pérot cavities and diameter for whispering-gallery-mode/spiral microresonators. All dots correspond to experimental results \cite{hirata2014sub, alnis2008subhertz, liu2024ultrastable, parke2026laser, guo2022chip, kelleher2023compact, mclemore2022miniaturizing, chen2014compact, hafner20158, ma2026laser, didier2018ultracompact, matei2017ultrastablelaser, chen2025laser, kedar2023frequency, liu202236, loh2025optical, li2026hertz, alnis2011thermal}. The figure also shows the noise limits of different systems: thermorefractive noise for spiral MRs (orange area) and thermal noise for FP cavities (red, purple, green, and blue lines). The noise limit of a spiral MR depends on its properties, including the effective mode volume and the material characteristics of the resonator (density, thermo-optic coefficient, thermal conductivity, and specific heat capacity) \cite{huang2019thermorefractive}. Therefore, the area between the orange lines represents the spiral MR noise limit for the most realistic range of parameters. The lines for FP cavities illustrate the dependence of thermal noise on the materials (spacer+coatings+mirrors): ULE — ultra-low-expansion glass, FS — fused silica, Si — monocrystalline silicon, "diel. mirr." — multilayer mirrors made of $\mathrm{SiO}_2/\mathrm{Ta}_2\mathrm{O}_5$, and "cryst. mirr." — mirrors made of GaAs/AlGaAs layers \cite{cole2013tenfold}. Silicon systems cannot be made compact because they require cooling to cryogenic temperatures.
    %\color{blue} References:
    %ULE+ULE+diel. mirr. -- \cite{hirata2014sub, alnis2008subhertz, liu2024ultrastable}; 
    %ULE+FS+diel. mirr. -- \cite{parke2026laser, guo2022chip, kelleher2023compact, mclemore2022miniaturizing, chen2014compact, hafner20158}; 
    %ULE+FS+cryst. mirr. -- \cite{ma2026laser, didier2018ultracompact}; 
    %Si+Si+diel. mirr. -- \cite{matei2017ultrastablelaser}; 
    %Si+Si+cryst. mirr. -- \cite{chen2025laser, kedar2023frequency}; 
    %Spiral MR -- \cite{liu202236, loh2025optical}; 
    %WGM MR -- \cite{li2026hertz, alnis2011thermal} \color{blue}
    }
    \label{discussion_picture}
\end{figure*}
One of the dominant limitations in this region is the relatively high intrinsic noise of the servo controller (see Appendix D). In the present work, we intentionally employ a simple digital controller to maintain a compact architecture and avoid complex electronic hardware. Nevertheless, modern digital servo controllers can simultaneously provide low input noise and low latency \cite{Liu2024control} and could be incorporated into future implementations.

A more challenging issue for compact Fabry--Pérot cavities is mitigating the effects of mechanical vibrations while avoiding bulky and complex active isolation systems. Our frequency noise power spectral density exhibits several pronounced vibration-induced peaks, indicating that vibration suppression will be essential for further performance improvements. One promising approach is to monitor the vibrations and compensate for their effect using a feed-forward correction scheme. 

Different applications sample different segments of the demonstrated noise spectrum (see Fig.~\ref{allan sil laser}). Optical clocks are mostly sensitive to laser instability at interrogation times of 0.1--10~s. The thermal-noise-limited performance of our system at 1--10~s directly supports clock instabilities at the low-$10^{-15}/\sqrt{\tau}$ level, comparable to state-of-the-art transportable clocks \cite{grotti2018geodesy,grotti2024long}. Such transportable clocks are the key instruments of chronometric leveling, where they resolve centimeter-scale geopotential differences in field campaigns. 
Microwave generation by optical frequency division \cite{kudelin2024photonic, ji2025dispersive,diakonov2025hybrid}, in contrast, relies on phase noise at Fourier frequencies of $\sim$1~kHz--1~MHz, which is important for radars and communications. Assuming ideal frequency division of the presented laser system down to 10~GHz (division factor $\sim$20 000), the optical reference would contribute a projected phase noise of less than -150~dBc/Hz at 10~kHz offset. As an important result, we demonstrate the PDH stabilization of the DFB laser diode serving as a comb pump tooth in the frequency comb generation regime, thus highlighting the potential of the designed system for stability transfer from optical into the microwave domain.

The place of this work among world results is shown in Fig.~\ref{discussion_picture}. This graph illustrates the theoretical instability limit of laser systems compared to experimental results obtained worldwide. The right part shows the results for laboratory-scale ultrastable laser systems with long reference Fabry-Perot cavities. The left part shows the systems of interest to us in this study, which have compact key components. Overall, the global community is interested in moving closer to the lower left corner of this graph. According to the picture, among these systems, the Allan deviation for the designed SIL laser is smaller than in other similar studies \cite{leibrandt2011field}. 

\section{Conclusion}
%ПИШЕМ ВЫВОДЫ ПРО КОМПАКТНЫЙ ЛАЗЕР НА SIL И МИНИ-ФП
%We make fantastic stable laser! =)

We have demonstrated an ultrastable laser system that combines a hybrid-integrated self-injection-locked diode laser based on a high-$Q$ Si$_3$N$_4$ ring microresonator with a miniature (3~ml) vacuum-gap Fabry--P\'erot reference cavity. Careful characterization and suppression of the technical noise of the electro-optical feedback loop — most notably, active stabilization of residual amplitude modulation, which reduced the long-term instability by more than an order of magnitude — allows the system to reach the fundamental Brownian thermal noise floor of the reference cavity. The fractional frequency instability of $5\cdot10^{-15}$ at averaging times of 1--10~s places this system among the most stable compact lasers reported to date, on the same level as results obtained with microfabricated mirrors of an order of magnitude higher finesse, while using conventional multilayer mirrors and a moderate finesse of $1\cdot10^{5}$. 
%In the Discussion section, we outline several straightforward upgrades that could further improve the system performance.Although the complete experimental setup is not yet highly compact, its core components—the self-injection-locked laser and the Fabry--P\'erot cavity-are sufficiently small to enable the development of significantly more compact devices, since further miniaturization of the vacuum and optical systems and electronics is primarily an engineering challenge. 
Furthermore, the designed experimental setup also supports Kerr frequency comb generation. Preliminary experiments indicate the feasibility of direct Pound--Drever--Hall stabilization of the comb pump tooth, thus highlighting the prospect of the stability transfer from optical into the microwave domains via optical frequency division techniques. The demonstrated architecture provides a pathway toward a fully integrated, thermal-noise-limited photonic frequency reference for transportable optical clocks and low-noise microwave synthesis.

\section{Methods}

\subsection{Hybrid integrated SIL laser}
Since the DFB diode has a linewidth of approximately 1 MHz, it is problematic to directly lock it to the FP cavity using the PDH technique \cite{Nakamura2025}, and preliminary linewidth narrowing by self-injection locking is therefore essential. We operate at the DFB current $I=215$ mA, ensuring the linear SIL regime and single-frequency operation for laser stabilization. 
We estimate the SIL laser linewidth (without locking to an external FP cavity) to be equal to 4 kHz, while the linewidth in the free-running regime is 900 kHz (these values are estimated by integrating the phase noise from high to low Fourier frequencies until the integrated phase noise exceeds $1 \ \mathrm{rad}^2$ \cite{matei2017ultrastablelaser}). The linewidth of the SIL laser is basically limited by the TRN of the MR material \cite{matsko2007whispering}. To estimate the long-term performance of the SIL laser without locking to an external reference, we perform measurements of the Allan deviation. The Allan deviation of the SIL laser without external FP-cavity stabilization is found to be $3\cdot 10^{-11}$ at an averaging time of 1 s (see Appendix G).

%After obtaining the optimal phase the frequency noise of the SILed laser comparing to the free-run changes almost $10^4$ times, see Fig.~\ref{free_run_vs_sil}. The instantenuous linewidth of the laser becomes close to the thermo-refractive noise \cite{huang2019thermorefractive}. 

%\begin{figure}[h]
%    \centering
%    \includegraphics[width=\linewidth]{figs/free_run_vs_sil_ylim_and_TRN.png}
%    \caption{Experimentally measured frequency noise PSD for the DFB laser in the free-running (without optical feedback from MR) regime and in the self-injection locking (SIL) regime. Gray dotted line shows the analytically calculated thermorefractive noise (TRN) level for the MR under study in the infinite heat bath approximation \cite{huang2019thermorefractive}.}
%    \label{free_run_vs_sil}
%\end{figure}

%In fact, the cutting-edge technologies imply the heterogeneous integration of the active medium on the same substrate as the passive MR. However, these technologies are not still widely available. Additionally, at the stage of research work the hybrid integration, in contrast to heterogenous one, allows to flexibly change the DFB laser diodes and photonic chips to find the best combination of their parameters. Thus, we use the hybrid integrated laser based on the self-injection locking of the DFB diode frequency to a microring resonator mode.

\subsection{Microresonator with normal dispersion}

%\begin{figure}[h]
%    \centering
%    \includegraphics[width=\linewidth]{figs/mr_4_panels.png}
%    \caption{(a) The photo of the photonic chip with the MR under study. (b) The resonant profile of the MR eigenmode at $\lambda=1545.0$ nm. (c) The photonic chip cross-section. The waveguide core is made of silicon nitride $\mathrm{Si_3N_4}$, while the cladding is made of  silicon dioxide $\mathrm{SiO_2}$. (d) The fundamental TE mode field distribution at $\lambda=1550.2$ nm.}
%    \label{chip_photo}
%\end{figure}

To demonstrate the long-term stability of the SIL laser in the single-frequency operation regime (Fig.~\ref{allan sil laser}), we use $\mathrm{Si_3N_4}$ MR with normal GVD, a diameter of $D=2.27$ mm, corresponding to a free spectral range of $\mathrm{FSR}=21.3$ GHz, with the relatively thin ring waveguide (thickness 300 nm, width 3 $\upmu$m). The MR under study features a high quality factor $Q=8.2\cdot10^6$ (average quality factor of TE eigenmodes within the wavelength range $1510-1630$ nm). Other parameters of the MR are presented in Table~\ref{mrr_params}. The characterization technique is described in Appendix B.
% The effective mode volume for the MR under study is equal to $V_\mathrm{eff}=22.8\cdot10^3$ $\mathrm{\upmu m^3}$.

\begin{table}[H]
  \centering
  \begin{tabular}{ccc}
    \hline
    \hline
    Parameter & Value & Unit\\
    \hline
    $\mathrm{Q}$ & $8.2\cdot10^6$ & - \\
    $\mathrm{FSR}$ & 21.3 & $\mathrm{GHz}$ \\
    $\mathrm{\beta_2}$ & 842.3 & $\mathrm{ps^2/km}$ \\
    $\kappa_0/2\pi$ & 16.1 & MHz \\ 
    $\kappa_c/2\pi$ & 7.6 & MHz\\
    $ g/2\pi$  & 6.7 & MHz \\
    \hline
    \hline
  \end{tabular}
  \caption{The results of the characterization of the MR with normal GVD. $Q$ is the MR quality factor, FSR -- free spectral range, $\beta_2$ -- group velocity dispersion coefficient, $\kappa_0$ and $\kappa_c$ are the intrinsic and coupling loss rates, respectively, $g$ is the backscattering coefficient. FSR and $\beta_2$ are determined at $\lambda=1545.0$ nm. Parameters $Q,\kappa_0, \kappa_c$ are averaged over all measured MR resonance profiles within the wavelength range 1510-1630 nm. }
  \label{mrr_params}
\end{table}

% \begin{table}[h]
%     \centering
%     \begin{tabular}{ccc}
%     \hline
%     \hline
%     Parameter & Value & Unit\\
%     \hline

%     \hline    
%     \hline
%     \end{tabular}
%     \caption{}
%     \label{dispersion params}
% \end{table}

\subsection{Microresonator with anomalous dispersion}

To examine the designed experimental system in a Kerr comb generation regime (Fig.~\ref{comb_and_freq_noise}), we employ another $\mathrm{Si_3N_4}$ MR with anomalous GVD, a diameter of $D=1.56$ mm, corresponding to a free spectral range of $\mathrm{FSR}=30.6$ GHz, and a quality factor of $Q=6.3\cdot10^6$ (the average quality factor of TE eigenmodes within the wavelength range $1510-1630$ nm).  Other parameters of the MR are presented in Table~\ref{mrr_anomalous_gvd_params}. The characterization technique is described in Appendix B.
% The effective mode volume for the MR under study is equal to $V_\mathrm{eff}=22.8\cdot10^3$ $\mathrm{\upmu m^3}$.

\begin{table}[H]
  \centering
  \begin{tabular}{ccc}
    \hline
    \hline
    Parameter & Value & Unit\\
    \hline
    $\mathrm{Q}$ & $6.3\cdot10^6$ & - \\
    $\mathrm{FSR}$ & 30.6 & $\mathrm{GHz}$ \\
    $\mathrm{\beta_2}$ & -142.7 & $\mathrm{ps^2/km}$ \\
    $\kappa_0/2\pi$ & 20.6 & MHz \\ 
    $\kappa_c/2\pi$ & 10.3 & MHz\\
    $ g/2\pi$  & 7.7 & MHz \\
    \hline
    \hline
  \end{tabular}
  \caption{The results of the characterization of the MR with anomalous GVD. $Q$ is a MR quality factor, FSR -- free spectral range, $\beta_2$ -- group velocity dispersion coefficient, $\kappa_0$ and $\kappa_c$ are intrinsic and coupling loss rates respectively, $g$ - backscattering coefficient. FSR and $\beta_2$ are determined at $\lambda=1567.0$ nm. Parameters $Q,\kappa_0, \kappa_c$ are averaged over all measured MR resonance profiles within the wavelength range 1510-1630 nm. }
  \label{mrr_anomalous_gvd_params}
\end{table}

\subsection{Fabry--Pérot cavity}
In this work, we use a Fabry--Pérot cavity with a volume of approximately 3 ml and a finesse of about $105 \ 000$. One of its mirrors is flat ($R_1=\infty$), and the other one is concave with a radius of curvature $R_2=23\,$cm. The cavity (both spacer and mirror substrates) is made of ultra-low-expansion (ULE) glass, whose coefficient of thermal expansion crosses zero near room temperature. The characteristics of the Fabry--Pérot cavity are provided in Table~\ref{fp_characteristics}

\begin{figure}[htbp!]
    \centering
   \includegraphics[width=0.7\linewidth]{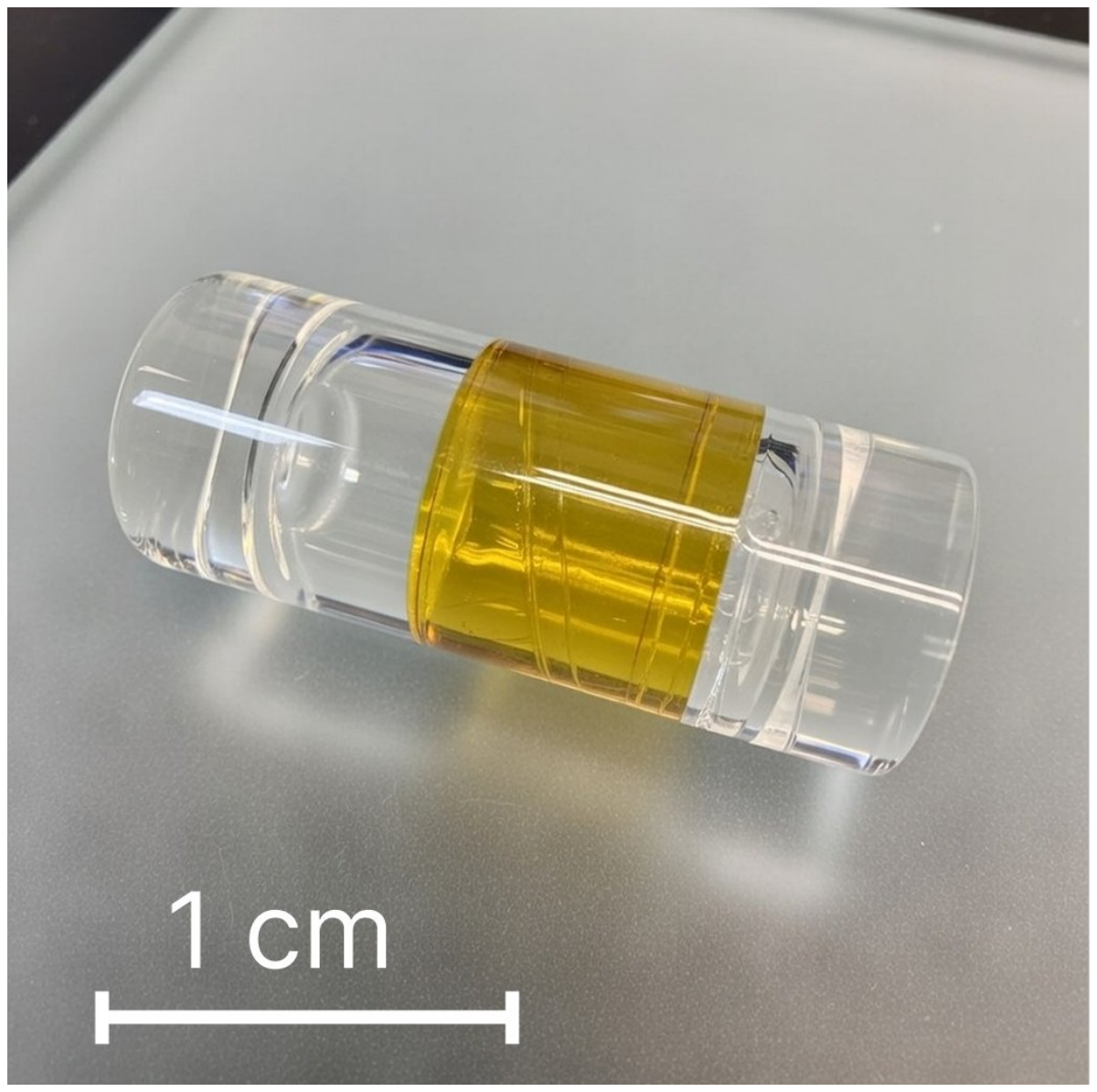}
   \caption{Photograph of the ULE Fabry--Pérot cavity.}
   \label{fp_photo}
\end{figure}

\begin{table}[h]
  \centering
  \begin{tabular}{ccc}
    \hline
    \hline
    Parameter & Value & Unit\\
    \hline
    $R_1$ & $\infty$ & $\mathrm{cm}$ \\
    $R_2$ & $23$ & $\mathrm{cm}$ \\
    $L$ & $2$ & $\mathrm{cm}$ \\
    $\mathcal{F}$ & 105 000 & - \\
    $Q$ & $2.7\cdot10^9$  & - \\
    $\mathrm{FSR}$ & 7.495  & $\mathrm{GHz}$\\
    $\delta\nu$ & 71 & $\mathrm{kHz}$ \\
    $R_{cav}$ & 27 & \% \\
    $T_{cav}$ & 28 & \% \\
    $T_0$ & 27.35 & $^\mathrm{o}\mathrm{C}$ \\
    \hline
    \hline
  \end{tabular}
  \caption{Characteristics of the Fabry--Pérot cavity. $R_1$, $R_2$ -- mirror radii of curvature, $L$ -- length of the cavity, $\mathcal{F}$ -- finesse. FSR (free spectral range) is the frequency spacing between two adjacent longitudinal modes, and $\delta \nu$ is the resonance linewidth of the FP cavity, $R_{cav}$ and $T_{cav}$ - reflection and transmission of the cavity at resonance, $T_0$ - zero-expansion temperature.}
  \label{fp_characteristics}
\end{table}

In the case of a perfect stabilization system, the limit of fractional frequency instability of a laser locked to the cavity is defined by fluctuations of the cavity length. The fundamental limit of instability to these fluctuations is set by the cavity thermal noise \cite{kessler2011fp_thermal_noise}. This noise originates from the thermal motion of the cavity spacer, mirror substrates, and coatings. For the FP cavity under study, we calculate the thermal noise limit of fractional frequency instability to be approximately $5\cdot10^{-15}$ (for the detailed analysis, see Appendix A).

To minimize the impact of environmental noise, we place the FP cavity into a vacuum chamber with a pressure below $10^{-8}$ mbar and employ active temperature stabilization at zero-expansion temperature $T=27.35$ $^\mathrm{o}\mathrm{C}$ with 1 mK stability. The chamber volume is less than $3\,$l. However, the present design includes redundant thermal shielding, and the chamber volume could be further reduced. The cavity is supported by three pairs of rods arranged 120$^\circ$ apart. The entire support structure was optimized using finite-element analysis to minimize its susceptibility to vibrations and is made of polyether ether ketone (PEEK) to reduce thermal coupling between the cavity and its environment. The developed support structure fixes the FP cavity in all directions and provides mechanical stability to tilting and acceleration. We measured the drift of the cavity resonance frequencies over a 24-hour period and found the drift rate to be less than 100 mHz/s.

\section{Acknowledgments}
%The experimental work is supported by Rosatom in the framework of the Roadmap for Quantum computing (Contract No. 868-1.3-15/15-2021 dated October 5).
The silicon nitride chips are fabricated by Qaleido Photonics. The work is supported by the Russian Science Foundation (Grant No. 26-72-10176).

\section{Data availability}
The data that support the findings of this study are
available from the authors upon reasonable request.

\newpage
\section{APPENDIX A: Calculation of Fabry--Pérot cavity thermal noise}
The standard approach to calculating cavity thermal noise \cite{kessler2011fp_thermal_noise} is to account for the contributions from the spacer ($S_\mathrm{sp}$), mirror substrates ($S_\mathrm{sub}$), and mirror coatings ($S_\mathrm{coat}$) using the expressions \eqref{eq_spacer}-\eqref{eq_coat} below. We denote $E$ and $\phi$ to be the Young's modulus and loss angle respectively. The quantities $S_\mathrm{sp}$, $S_\mathrm{sub}$ and $S_\mathrm{coat}$ have the dimension of length squared per unit frequency.

The cavity spacer contribution is given by \cite{kessler2011fp_thermal_noise}:

\begin{equation}
S_\mathrm{sp}(f)=\frac{2k_BT}{\pi^2f}\cdot \frac{L}{R^2-r^2} \cdot \frac{\phi_\mathrm{sp}}{E_\mathrm{sp}},
\label{eq_spacer}
\end{equation}
where $f$ is the Fourier frequency, $L$ is the cavity length, $R$ and $r$ are the outer and inner radii, respectively, of the cavity spacer, $T$ is the cavity temperature, and $k_B$ is the Boltzmann constant. 

The corresponding expression for cavity substrate contribution reads:

\begin{equation}
S_\mathrm{sub}(f)=\frac{2k_BT}{\pi^{3/2}f}\cdot \frac{1-\sigma_\mathrm{sub}^2}{w} \cdot \frac{\phi_\mathrm{sub}}{E_\mathrm{sub}},
\label{eq_sub}
\end{equation}
where $\sigma_\mathrm{sub}$ is the Poisson's ratio of the mirror substrate and $w$ is the radius of the laser beam on the mirror.

The contribution from cavity coatings is given by

\begin{equation}
S_\mathrm{coat}(f)=S_\mathrm{sub}(f) \frac{2}{\sqrt{\pi}} \frac{1-2\sigma_\mathrm{sub}}{1-\sigma_\mathrm{sub}} \frac{\phi_\mathrm{coat}}{\phi_\mathrm{sub}} \frac{d_\mathrm{coat}}{w},
\label{eq_coat}
\end{equation}
where $d_\mathrm{coat}$ is the thickness of the mirror coating.

%\begin{figure}[H] 
%    \centering
%    \includegraphics[width=0.25\linewidth]{figs/cavity_noise.png}
%    \caption{Three contributions to total cavity thermal noise. 1 -- substrate, 2 -- mirror coatings, 3 -- spacer}
%    \label{fp_cavity_noise}
%\end{figure}

The total thermal noise is given by
\begin{multline}\label{fp_total_noise}
S_\mathrm{tot}(f) = S_\mathrm{sp}(f) + S_\mathrm{sub1}(f) + \\S_\mathrm{sub2}(f) + S_\mathrm{coat1}(f) + S_\mathrm{coat2}(f),
\end{multline}
where the system has two mirror substrates and two mirror coatings. The calculated contributions from the different parts of the cavity used in this work are listed in Table~\ref{thermal noise}.

\begin{table}[H]
  \centering
  \begin{tabular}{ccc}
    \hline
    \hline
    Type & PSD ($\mathrm{m}^2/f$) \\
    \hline
    $S_\mathrm{sub}$ & $2.4 \cdot 10^{-33}/f$ \\
     & $2.3 \cdot 10^{-33}/f$ \\
    $S_\mathrm{coat}$ & $1.7 \cdot 10^{-33}/f$ \\
     & $1.5 \cdot 10^{-33}/f$ \\
    $S_\mathrm{sp}$ & $2.0 \cdot 10^{-34}/f$ \\
    $S_\mathrm{tot}$ & $8.1 \cdot 10^{-33}/f$ \\
    \hline
    \hline
  \end{tabular}
  \caption{Results of the thermal noise calculation for the cavity used in this work. For the mirror substrates and coatings, the first value corresponds to the flat mirror, while the second corresponds to the spherical mirror.}
  \label{thermal noise}
\end{table}

The fluctuations in the time domain are characterized by the Allan deviation $\sigma_y$ \cite{allan1981}. Thermal noise with a $1/f$ (flicker-frequency) power spectral density results in an Allan deviation that is independent of the averaging time:

\begin{equation}
\sigma_y = \dfrac{\sqrt{2\ln(2)S_\mathrm{tot}(f)f}}{L}.
\end{equation}

For the cavity used in this work, we calculate the fundamental limit on the laser fractional frequency instability to be:
\begin{eqnarray}
    \sigma_y =5\cdot10^{-15}.
\end{eqnarray}

\section{APPENDIX B: Microresonator characterization procedure}
\begin{figure}[h]
    \centering
    \includegraphics[width=\linewidth]{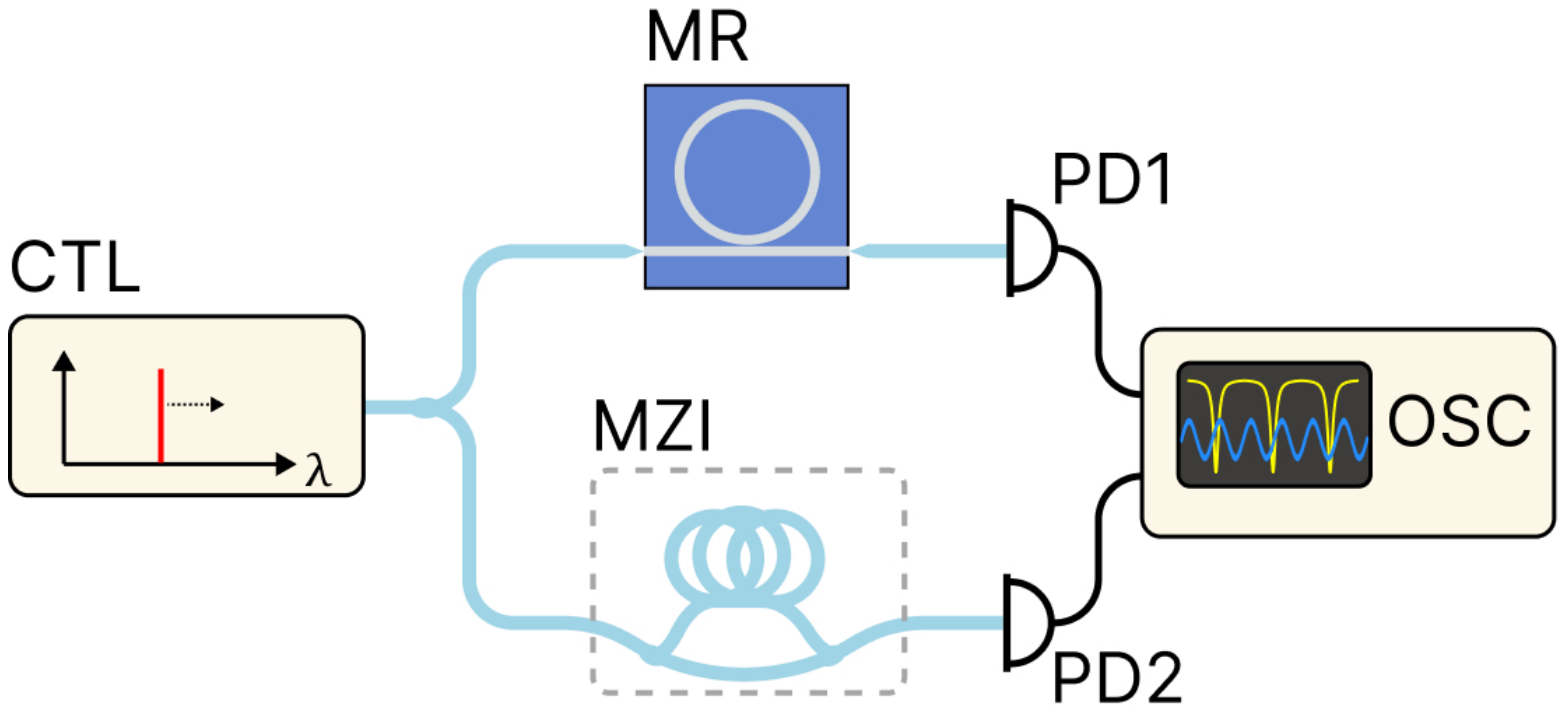}
    \caption{The experimental setup for MR  CTL - continuously tunable single frequency laser, MR - microring resonator under study, MZI - Mach-Zehnder interferometer, PD - photodetector, OSC - oscilloscope.}
    \label{dispersion data}
\end{figure}

For the measurement of MR parameters (dispersion profile, intrinsic and coupling loss rates, quality factors of eigenmodes), we measure the transmission spectrum of a MR under study. The characterization setup is shown in Fig.~\ref{dispersion data}. We use a continuously tunable laser (CTL) to record the transmission spectrum of a MR in the wavelength range 1510-1630 nm. We employ an auxiliary calibrated Mach-Zehnder interferometer (MZI) as a frequency reference. During a continuous wavelength scan of the laser source in the full available range, transmission spectra of both the investigated MR and the reference MZI are simultaneously recorded in the memory of the oscilloscope. Next, a recorded MZI transmission spectrum is used as a calibrated frequency scale. As a result, we obtain a precise MR transmission spectrum.

%In the measurements, we use the polarization, which is selected for the experimental workflow in this work [horizontal with respect to the plane of a photonic chip]. 

Based on the recorded transmission spectrum of a MR, we define intrinsic and coupling losses for eigenmodes of a MR under study. To extract parameters of measured resonance profiles, we employ a split Lorentzian fit, which takes into account the presence of a backscattered wave in a MR \cite{gorodetsky2000rayleigh, kondratiev2020modulational}.

Next, using the MR spectrum, we identify the MR eigenfrequencies $\omega_m$. In fact, the dispersion leads to the deviation of eigenfrequencies $\omega_m$ from the equidistant frequency grid. This may be expressed in the following form:
\begin{eqnarray}
    \omega_m=\omega_0+D_1m+\dfrac{1}{2}D_2m^2+\dfrac{1}{6}D_3m^3+...,
    \label{omega_mu}
\end{eqnarray}
where $\omega_0$ is a central frequency, $D_1/(2\pi)=\mathrm{FSR}$ and $D_k,\ k\geq2$ are the dispersion coefficients. 
It is convenient to omit the linear part in Eq.~\eqref{omega_mu} and use the integrated dispersion profile defined as follows:
\begin{eqnarray}
    D_\mathrm{int}(m) = \dfrac{1}{2}D_2m^2 + \dfrac{1}{6}D_3m^3 + ... 
    \label{profile}
\end{eqnarray}
Importantly, the second-order dispersion coefficient in Eq.~\eqref{profile} is proportional to the group velocity dispersion coefficient \cite{fujii2020dispersion}:
\begin{eqnarray}
    D_2 = -v_gD_1^2\beta_2,
    \label{D2_beta_2}
\end{eqnarray}
where $v_g$ is a group velocity, $\beta_2$ is a group velocity dispersion (GVD) coefficient, $D_2 < 0$ corresponds to normal GVD and vice versa.

%\section{APPENDIX: Allan deviation for SIL laser}

%To assess the effectiveness of self-injection locking, we measure the frequency noise of the free-running DFB laser diode and the DFB laser locked to the MR (Fig.~\ref{freq noise free running vs sil}). 
%The frequency noise in the SIL regime is several orders of magnitude lower across the entire measured frequency range. 

\section{APPENDIX C: Three-cornered hat method}
For the characterization of the individual frequency instability of the SIL laser, we compare it with 871-nm and 1140-nm lasers, each stabilized to 8-cm-long vertical ULE cavities, using a fiber optical frequency comb and the three-cornered hat method.

\begin{figure}[h]
    \centering
    \includegraphics[width=\linewidth]{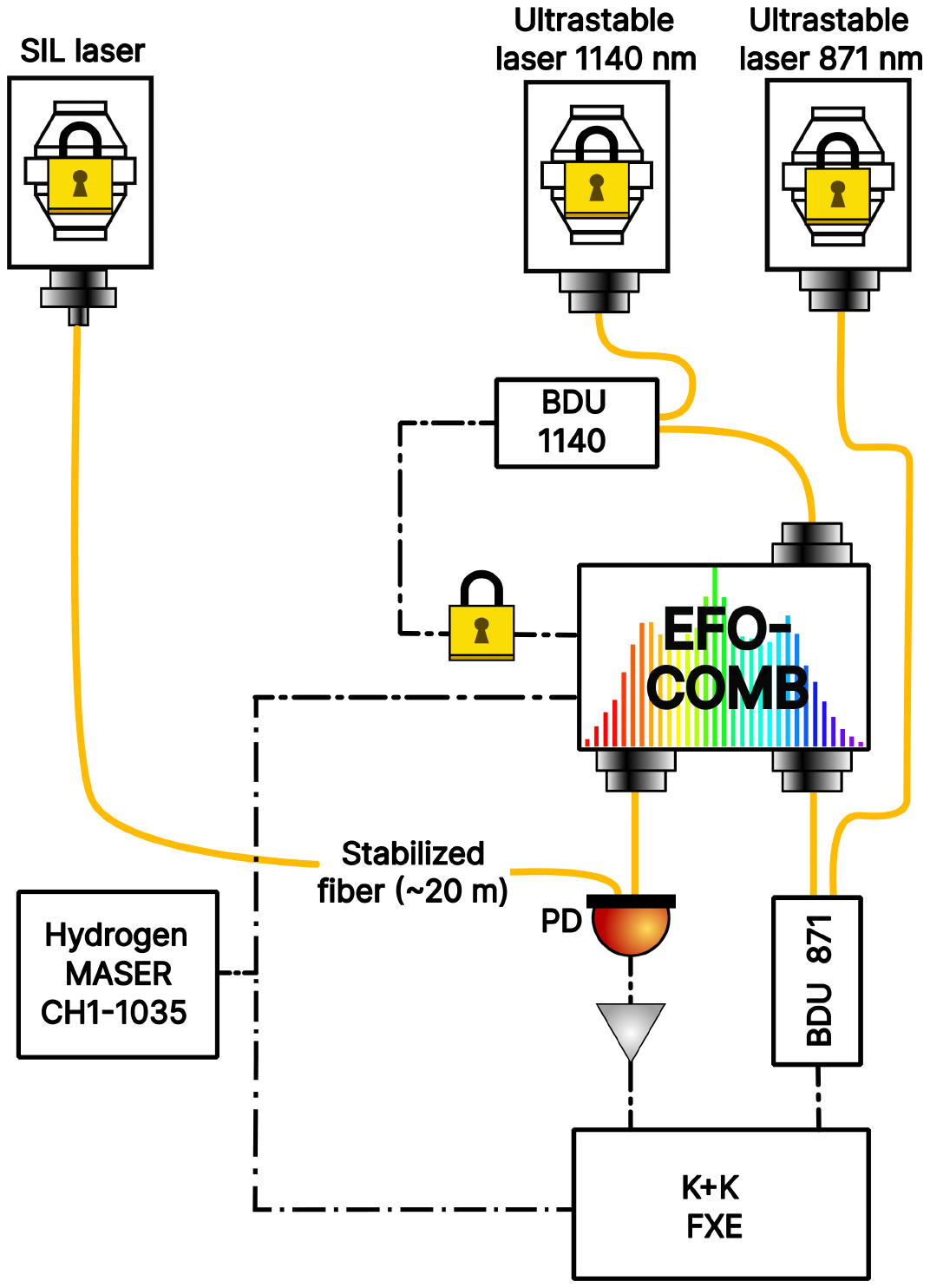}
    \caption{Measurement setup for the three-cornered hat method. The beats between the 1140-nm and 871-nm lasers and the optical frequency comb are measured using beat detection units (BDUs) based on balanced photodetectors, while the beat between the SIL laser and the comb is detected using a fast biased photodetector. The optical frequency comb is stabilized to the 1140-nm clock laser; at long averaging times, its frequency is additionally referenced to an RF signal from a hydrogen maser. The beat frequencies are recorded using a dead-time-free counter operating in $\Pi$-mode referenced by the same maser signal.}
    \label{three cornered hat scheme}
\end{figure}

The three-cornered hat method normally requires three lasers and the measurement of the beat frequency between each pair of lasers. Our experimental implementation differs slightly from this standard approach \cite{Kudeyarov2021}. As shown in Fig.~\ref{three cornered hat scheme}, we simultaneously record only two beat signals: the beat between the SIL laser and the frequency comb, and the beat between the 871-nm laser and the frequency comb. The fiber comb is stabilized to the 1140-nm laser, so these measurements provide the relative instabilities of the \textit{SIL+1140} and \textit{871+1140} beats, respectively. Hereafter, we refer to each laser system by its wavelength and denote the self-injection-locked laser simply as the \textit{SIL} system.

The third pairwise beat (\textit{SIL+871}) can be reconstructed from the two simultaneously recorded beat signals. Specifically, the corresponding frequency fluctuations are obtained by adding or subtracting the two measured signals, with the sign determined by the relative positions of the two laser frequencies with respect to the comb tooth, i.e., whether each laser frequency is higher or lower than the comb tooth frequency. Simultaneous acquisition of the \textit{SIL+1140} and \textit{871+1140} beats is essential for this reconstruction.

\begin{figure}[H]
    \centering
    \includegraphics[width=\linewidth]{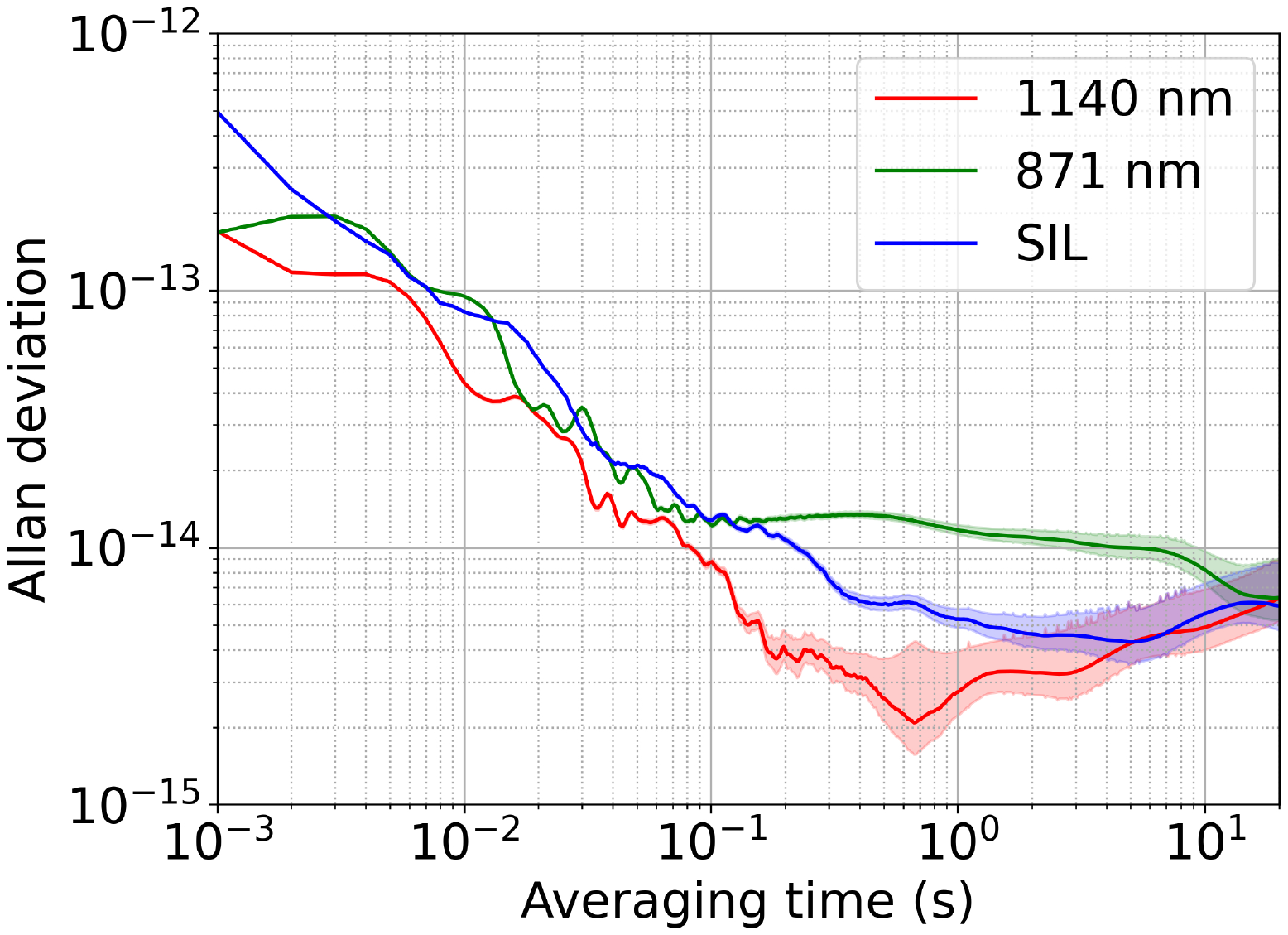}
    \caption{Fractional frequency instability of three lasers used in the three-cornered hat method. Errors are shown in translucent color}
    \label{three cornered hat results}
\end{figure}

Having obtained the relative frequency fluctuations for all three pairs—\textit{SIL+1140}, \textit{871+1140}, and \textit{SIL+871}—we can apply the three-cornered hat method to determine the individual frequency instabilities of the three laser systems. For the method to provide reliable results, the fractional instabilities of the systems should be of comparable magnitude; otherwise, the instability of the least stable system can dominate the pairwise measurements and lead to large uncertainties in the extracted instabilities. In our experiment, this condition is satisfied, and the resulting fractional frequency instabilities are shown in Fig.~\ref{three cornered hat results}. The error bars were calculated following the method described in \cite{riley2008handbook}.

\section{APPENDIX D: Characterization of feedback loop in PDH method}

Feedback noise in the PDH stabilization system can be divided into three components, each originating from a different part of the system and requiring a different measurement procedure: electronic noise from the detection circuit upstream of the servo controller, residual amplitude modulation (RAM) noise arising in the optical circuit, and noise introduced by the servo controller itself.

By electronic noise, we refer to the noise generated by the various electronic components of the detection circuit, including the photodetector, RF mixer (phase detector), and amplifiers.

Residual amplitude modulation (RAM) results from non-ideal phase modulation of the laser beam \cite{shen2015systematic, whittaker1985residual}. This non-ideality can arise from several effects, including temperature-dependent birefringence in the electro-optic modulator (EOM), the presence of second and higher-order harmonics during phase modulation, and parasitic reflections \cite{shen2015systematic}. For frequency stabilization, the absolute level of RAM is less important than its fluctuations over time, which are converted into frequency fluctuations by the PDH locking system.

For the servo controller, the primary concern is its input noise. Here, input noise refers to noise introduced into the signal before it is compared with the setpoint. Both input noise and fluctuations of the setpoint are indistinguishable from a signal caused by laser frequency fluctuations and therefore directly affect the frequency stability of the locked laser.

To identify the dominant noise source, we measure the contribution of each component to the frequency instability. For this purpose, a Rohde\&Schwarz HMC8012 precision digital multimeter was used to measure fluctuations of the error signal, i.e., the output voltage of the RF mixer used for PDH (see the schematic in Fig.~\ref{pdh_exp_setup}), under different operating conditions of the stabilization system.

\begin{enumerate}
\item \textbf{Electronic noise.} The laser radiation is blocked, allowing us to isolate the noise originating from the electronic components of the detection circuit without contributions from the optical signal.

\item \textbf{RAM noise:} The laser radiation is present, but the frequency stabilization loop is disabled, and the laser frequency is detuned from the cavity resonance by several MHz. Under these conditions, fluctuations of the demodulated signal originate from the optical path and its associated RAM rather than from fluctuations of the cavity resonance. Note that the noise obtained from this measurement also includes electronic noise.

\item \textbf{Servo controller noise.} The feedback loop is operated in the locked regime. The measured error signal fluctuations then characterize the residual noise that cannot be suppressed by the servo controller.
\end{enumerate}

The conversion of error-signal voltage fluctuations into frequency fluctuations is determined by the slope of the PDH error signal. Near the frequency $\nu_0$ corresponding to the maximum cavity transmission, the error signal can be approximated as a linear function of frequency \cite{black2001introduction}:
\begin{equation}
    U = D_\nu (\nu-\nu_0),
\end{equation}
where $U$ is the error signal voltage and $D_\nu$ is the slope of the error signal.

\begin{figure}[H]
    \centering
    \includegraphics[width=\linewidth]{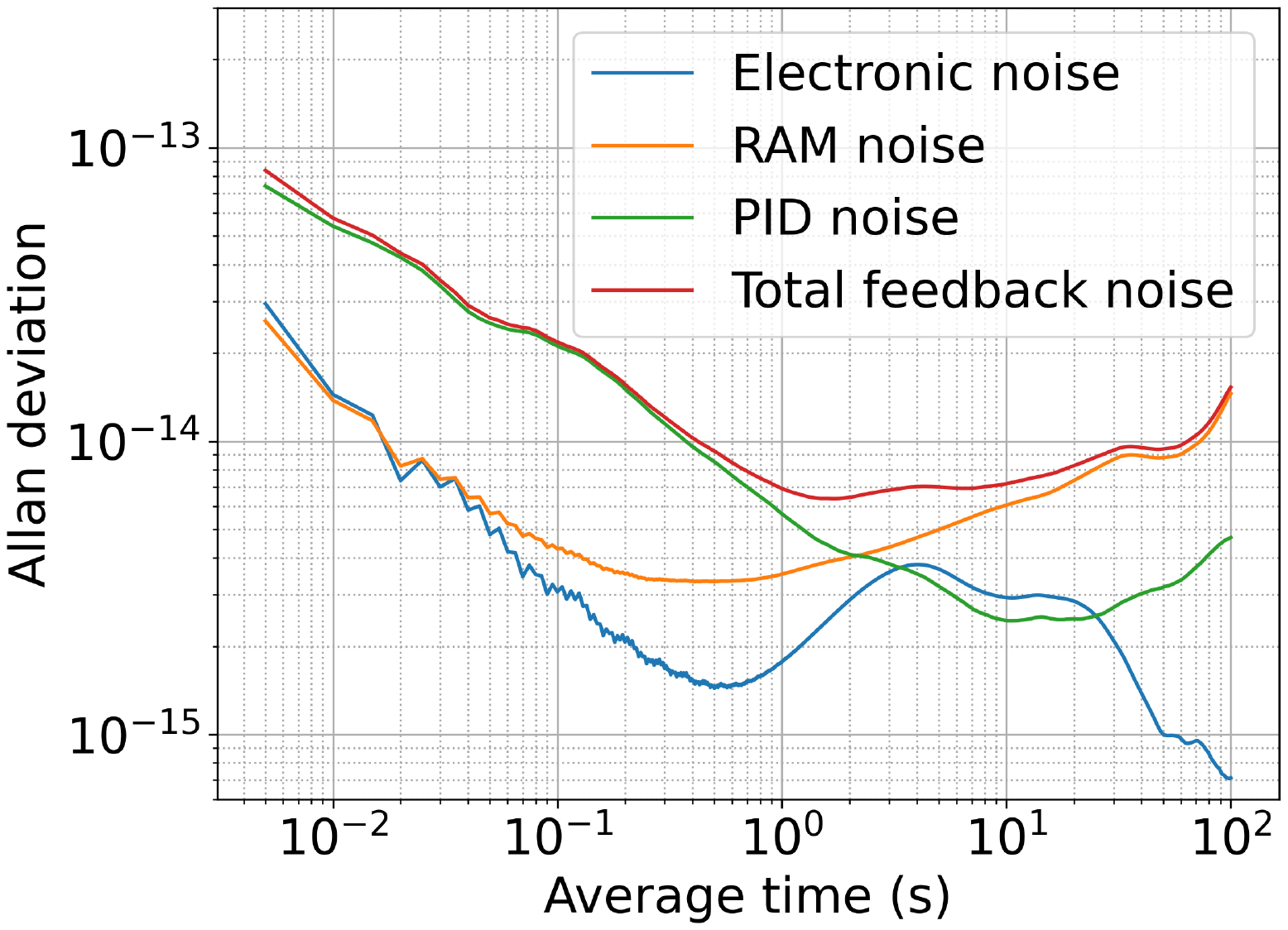}
    \caption{Example of measuring three types of feedback noise: electronic, RAM, servo controller; and the sum of these types of noise. From this picture, one can understand that there is a huge contribution of PID noise on small averaging times, while long-term stability of the feedback loop is defined by RAM fluctuations.}
    \label{feedback noise}
\end{figure}

We obtained that the error-signal slope is $1.7\cdot10^{-5}\,\mathrm{V/Hz}$. Dividing the measured error-signal fluctuations, expressed in volts, by this slope gives the corresponding frequency fluctuations. These can then be used to determine the fractional frequency instability associated with each noise type. An example of the measurements of the three noise contributions in our system is shown in Fig.~\ref{feedback noise}.

This procedure allows us to identify the dominant noise sources and determine which contributions should be suppressed first. In our system, the measurements reveal that servo-controller noise is the dominant limitation at short averaging times. The improvement can be achieved by using a servo controller with lower noise. RAM fluctuations become the main limitation at longer averaging times. We note that the combined contributions of feedback-related noise [Fig.~\ref{feedback noise}] and thermal noise slightly exceed the directly measured laser frequency instability [Fig.~\ref{allan sil laser}]. We attribute this discrepancy to the fact that the method is intended to identify the dominant noise sources rather than provide a precise quantitative estimate of the overall instability. Its accuracy may be limited by several factors: (1) some noise sources, such as residual amplitude modulation (RAM), cannot be monitored simultaneously with frequency fluctuations, meaning that certain experimental parameters, such as optical power, may differ slightly between measurements; and (2) dead time and nonuniform sampling in the data acquired using a multimeter can lead to biased estimates of the Allan deviation.

\section{APPENDIX E: Suppression of RAM}

\begin{figure}[h]
    \centering
    \includegraphics[width=\linewidth]{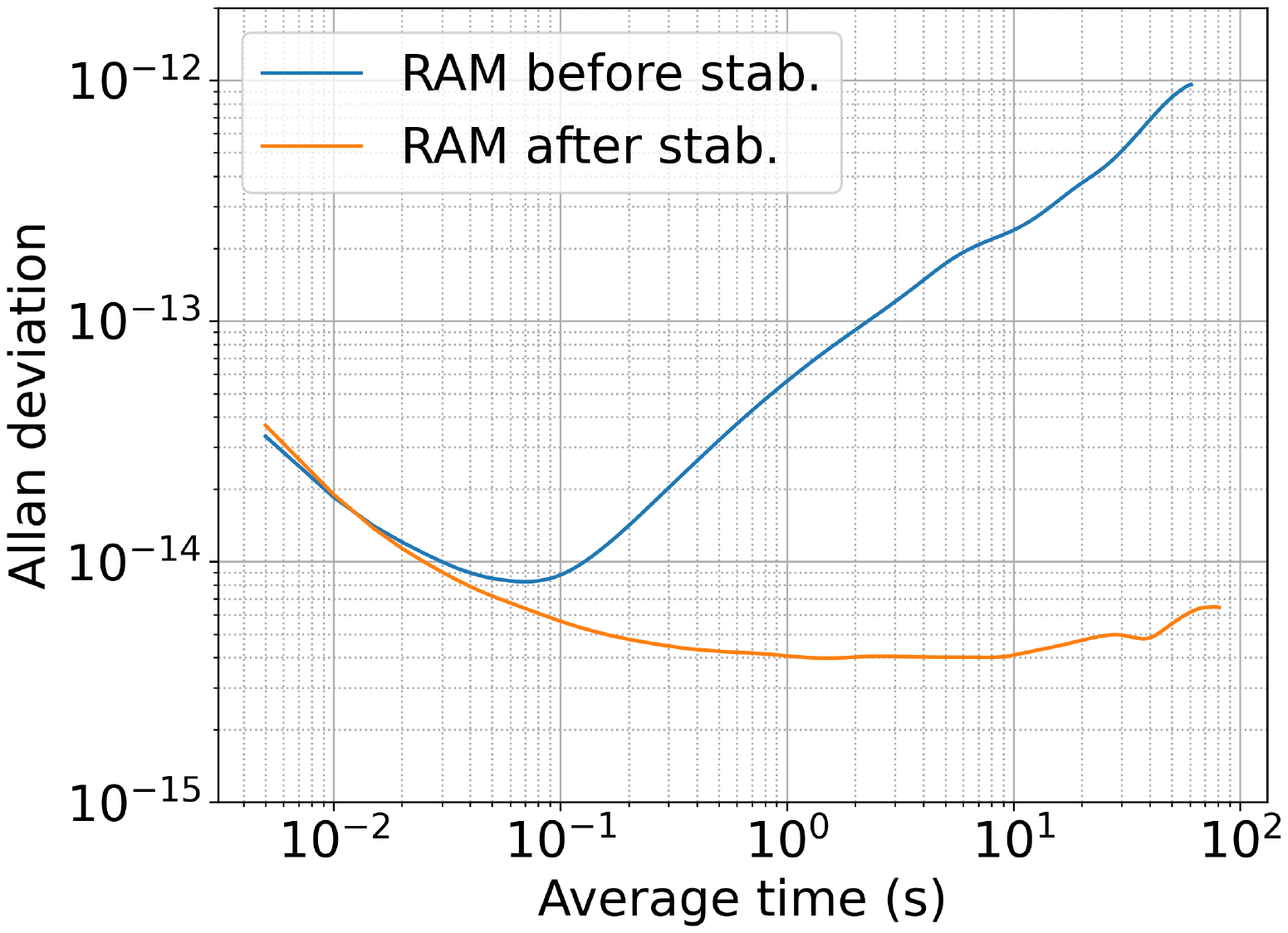}
    \caption{Fractional frequency instability induced by residual amplitude modulation (RAM) fluctuations before stabilization (blue) and with passive and active stabilization (orange). At averaging times greater than 1 s, RAM stabilization reduces the noise level by more than an order of magnitude.}
    \label{ram noise}
\end{figure}

The largest improvement in the feedback-loop noise was achieved through the suppression of RAM, using both active and passive techniques. As shown in Fig.~\ref{ram noise}, at averaging times greater than 1 s, RAM stabilization reduces the noise by more than an order of magnitude, enabling a fractional frequency instability below $5\cdot10^{-15}$.

\section{APPENDIX F: Linewidth calculation}

To carefully assess the performance of the stabilized laser and facilitate comparison with previous results, we estimate its optical linewidth using three different methods: the $\beta$-separation-line method, the $1\,\mathrm{rad}^2$ integrated phase-noise method, and the full width at half maximum (FWHM) of the beatnote signal measured with a spectrum analyzer. %The results are summarized in Table~\ref{linewidth}.

\begin{enumerate}
    \item The beatnote spectrum was directly recorded using a spectrum analyzer with a resolution bandwidth (RBW) of 3 Hz. From these data, we estimate the FWHM to be 4 Hz, which is probably mostly limited by the instrument RBW.
    
    %\begin{figure}[h]
    
    %\centering
    %\includegraphics[width=\linewidth]{figs/fwhm.png}
    %\caption{SIL laser beatnote with the frequency comb stabilized to the reference ultrastable laser.}
    %\label{lorenztian_fit}
    %\end{figure}
    
    \item The $\beta$-separation line method divides the frequency noise spectrum into two regions that affect the laser line shape in different ways. If $S_{\mathcal{\nu}} > f \cdot 8\ln 2/\pi^2$, the noise contributes to the laser linewidth; otherwise, it contributes only to the wings of the line shape. The line separating these two regions is called the $\beta$-separation line.

    The laser linewidth is calculated using the formula
    $$
    \Delta \nu_{\beta} = (8\ln 2 \cdot A)^{1/2},
    $$
    where $A$ is the integral of the frequency noise from the inverse of the measurement time up to the intersection between the frequency noise spectrum and the $\beta$-separation line.

    In Fig.~\ref{b line}, the frequency noise spectral density of the SIL laser studied in this work is shown together with the $\beta$-separation line. This method is not particularly suitable for ultrastable lasers, since most of their frequency noise lies below the $\beta$-separation line.

    \begin{figure}[H]
    \centering
    \includegraphics[width=\linewidth]{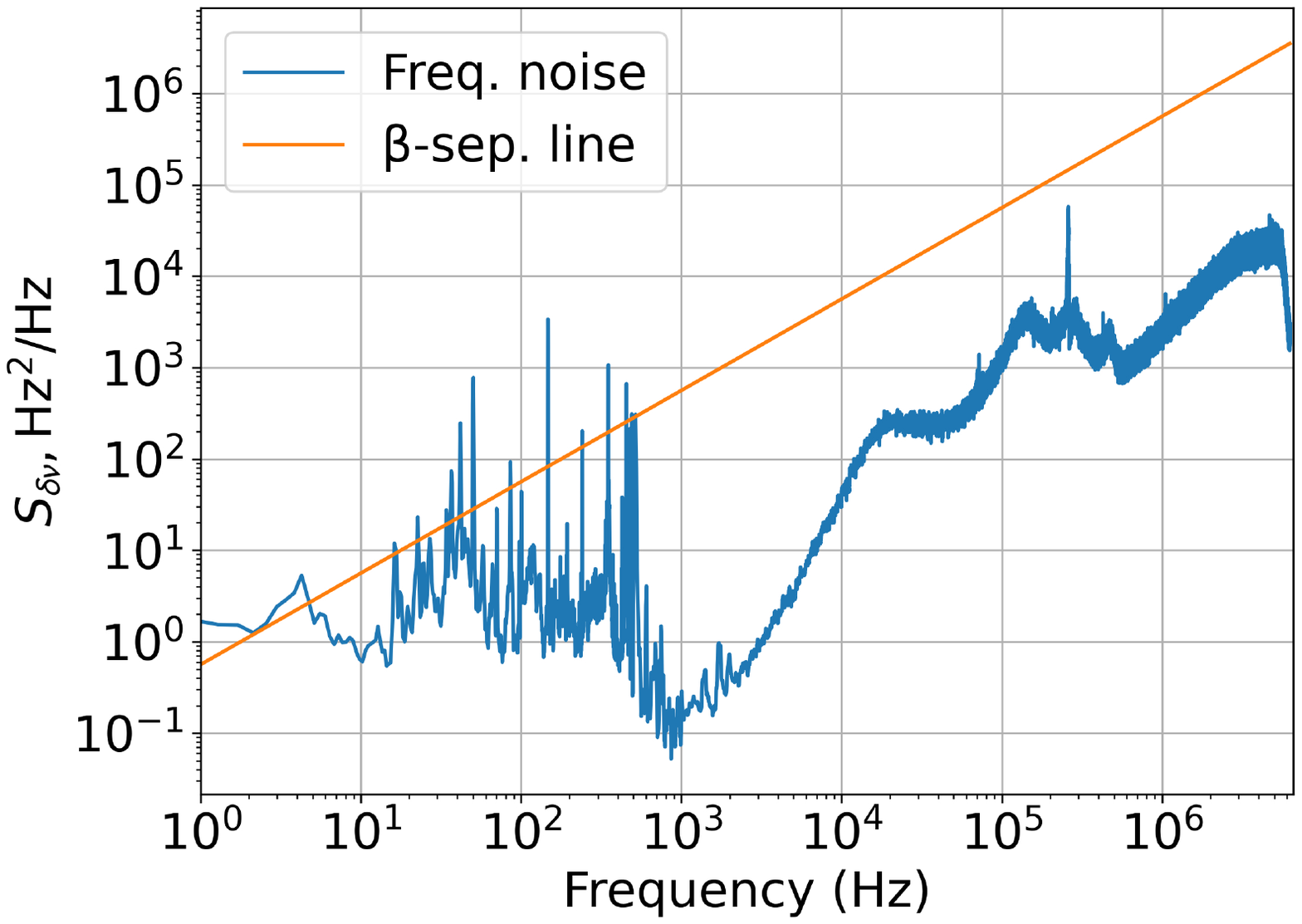}
    \caption{Frequency noise power spectral density and $\beta$-separation line of SIL laser. }
    \label{b line}
    \end{figure}

   \item In the $1$ rad$^2$-integral method, it is necessary to calculate the integral of the phase noise, starting from high frequencies and moving towards low frequencies. The frequency at which the integral exceeds $1 \  \mathrm{rad}^2$ is half the linewidth. Results of applying this method to the SIL laser are located in Fig.~\ref{1 pi linewidth}.

%по уровню 1 получилось 5.5 * 2 = 11 Гц, а по 1/pi rad^2 получилось 50 * 2 = 100 Гц

   \begin{figure}[H]
    \centering
    \includegraphics[width=\linewidth]{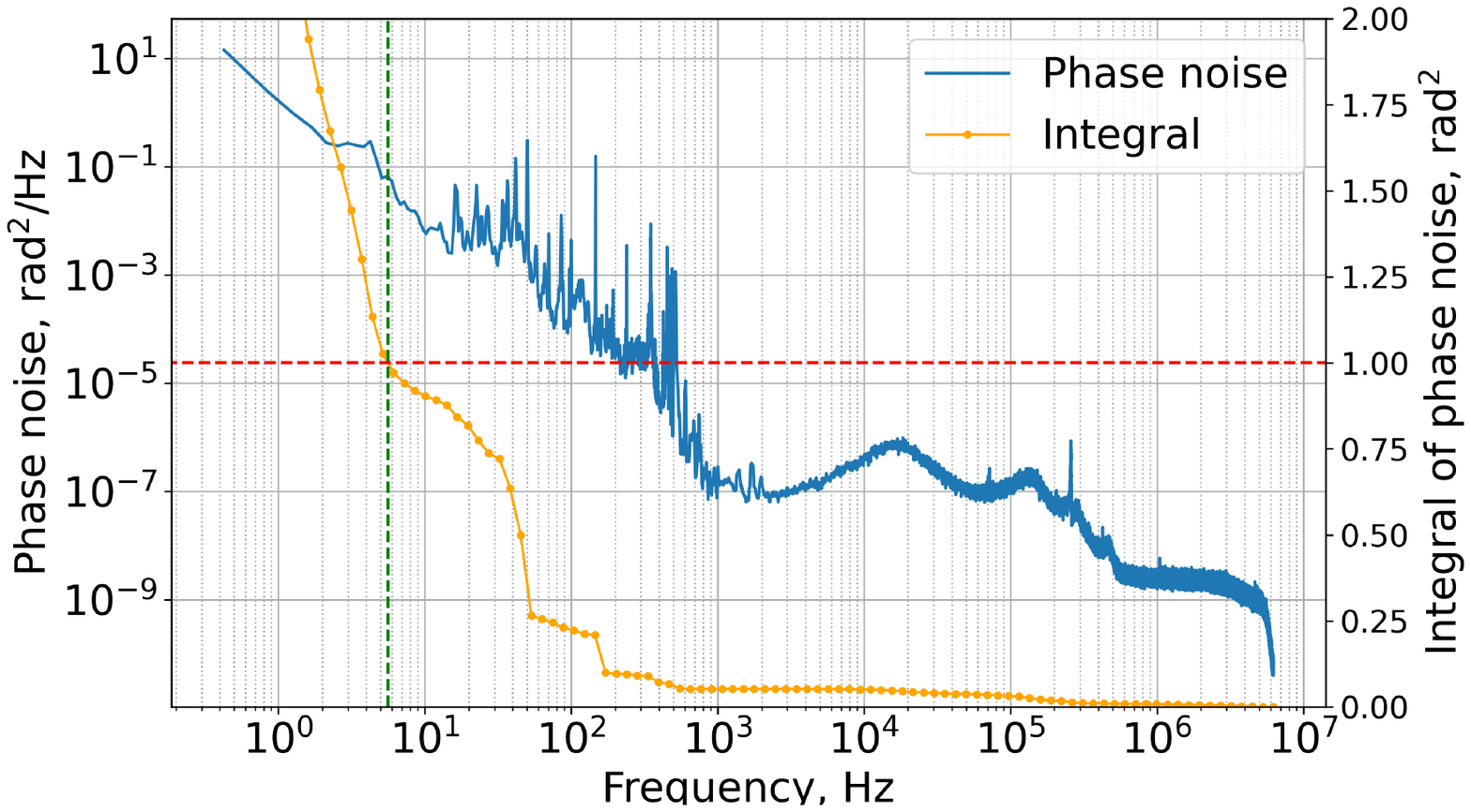}
    \caption{Phase noise power spectral density. The integral of phase noise to the right of the green line is equal to 1 rad$^2$}
    \label{1 pi linewidth}
    \end{figure}

\end{enumerate}

\begin{table}[h]
    \centering
    \begin{tabular}{cc}
        \hline
        \hline
        Method & Linewidth (Hz)\\
        \hline
        FWHM of beatnote, RBW=3 Hz & 4.0\\
        $\beta$-separation line, 1 s measurement time & 3.2\\
        $1$ rad$^2$-integral & 11\\
        \hline
        \hline
    \end{tabular}
    \caption{Laser optical linewidth estimations obtained by different methods. }
    \label{linewidth}
\end{table}

\section{APPENDIX G: Allan deviation of the SIL laser without PDH locking}
\begin{figure}[H]
    \centering
    \includegraphics[width=0.9\linewidth]{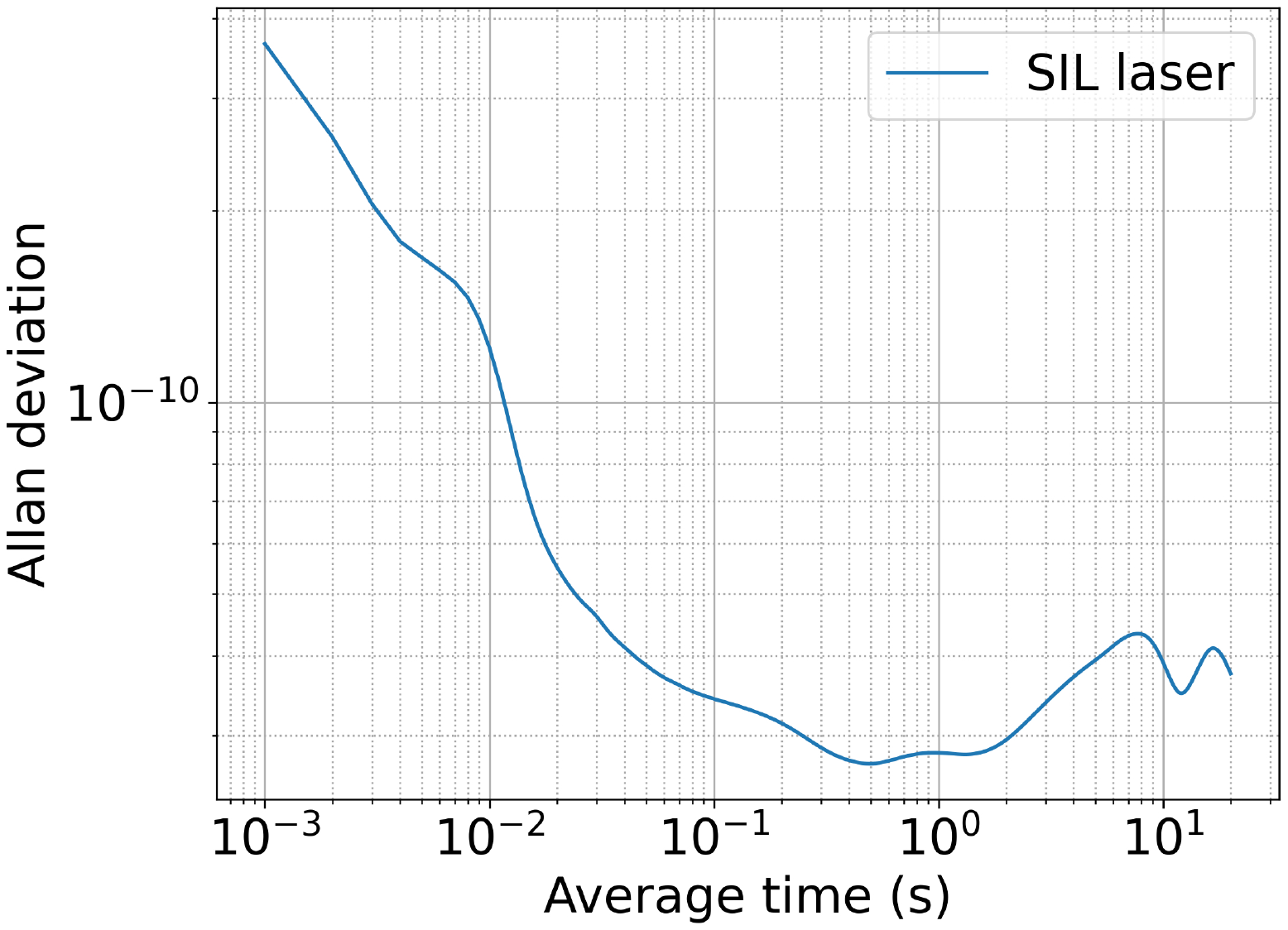}
    \caption{Fractional frequency instability of the SIL laser without PDH locking to the reference FP cavity.}
    \label{sil laser noise no pdh}
\end{figure}

Using the beatnote with the referenced Avesta frequency comb, we measured the Allan deviation of the laser operating in the SIL regime (Fig.~\ref{sil laser noise no pdh}), which is $3\cdot10^{-11}$ at an averaging time of 1 s.

%\begin{figure}[H]
%    \centering
%    \includegraphics[width=\linewidth]{figs/сравнение free running и sil.jpg}
%    \caption{Frequency noise PSD for laser in free-running (blue) and SIL (red) regimes}
%    \label{freq noise free running vs sil}
%\end{figure}

\section{APPENDIX H: Platicon microcomb}

\begin{figure}[h]
    \centering
    \includegraphics[width=0.8\linewidth]{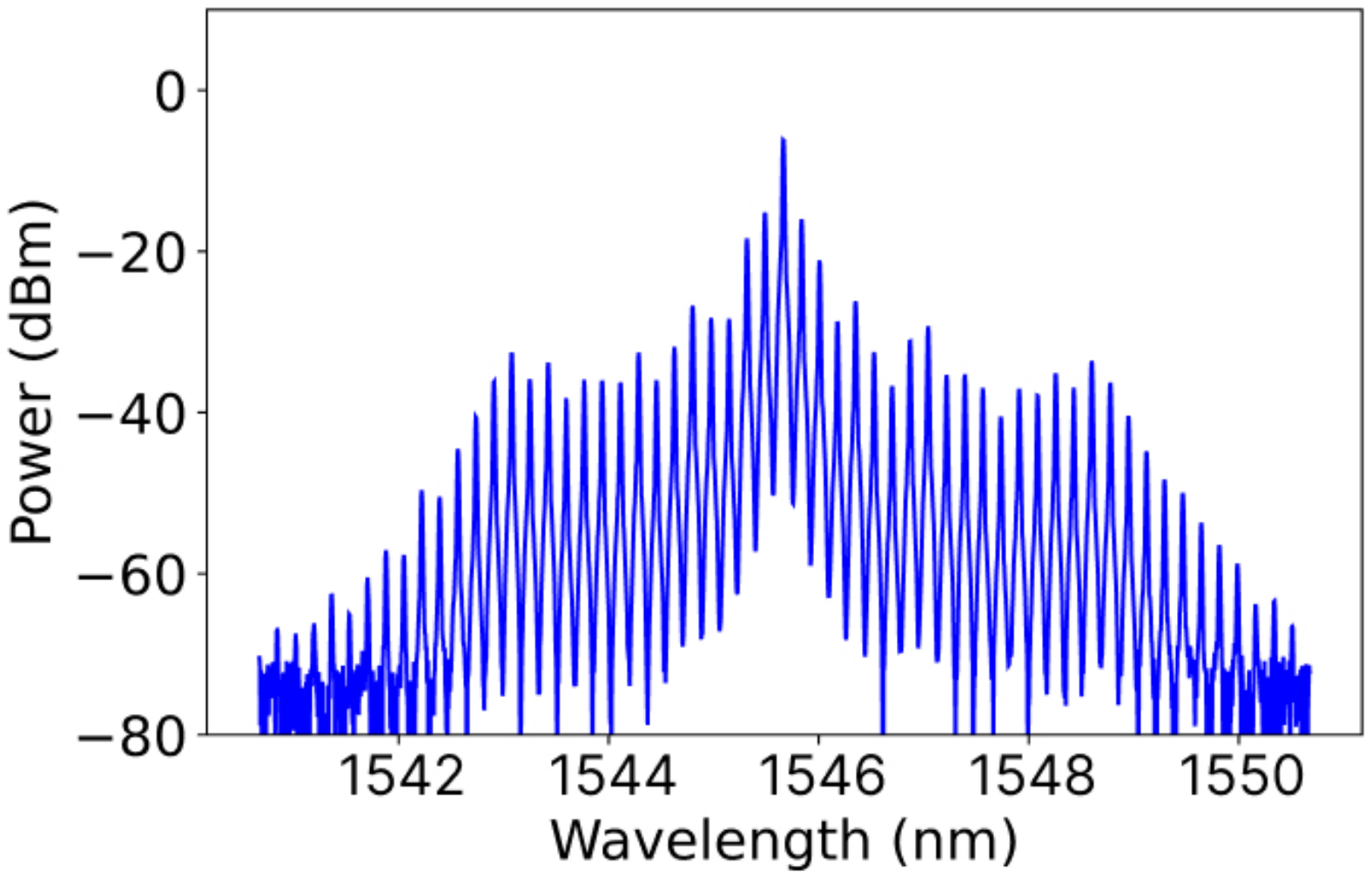}
    \caption{The spectrum of the platicon microcomb excited in the microresonator with normal group velocity dispersion in the self-injection locking regime.}
    \label{osa_spectrum}
\end{figure}

\newpage
\bibliography{lib}% Produces the bibliography via BibTeX.
\end{document}